\documentclass[11pt]{article}

\usepackage[utf8]{inputenc}
\usepackage[T1]{fontenc}
\usepackage{amsmath,amssymb}
\usepackage{microtype}
\usepackage[margin=1in]{geometry}
\usepackage{hyphenat}
\usepackage{silence}
\usepackage{graphicx}
\usepackage{booktabs}
\usepackage{multirow}
\usepackage{tabularx}
\usepackage{array}
\newcolumntype{C}{>{\centering\arraybackslash}X}
\usepackage{comment}
\usepackage[numbers,sort&compress]{natbib}
\usepackage{algorithm}
\usepackage{algpseudocode}
\usepackage{amsmath}
\usepackage{threeparttable}
\graphicspath{{figures/}}
\usepackage{enumitem}

\usepackage[
    bookmarks=true,
    colorlinks=true,
    linkcolor=blue,
    citecolor=red,
    urlcolor=blue,
    backref=page,
    pdftitle={Axion: A differentiable, GPU-native finite element framework, demonstrated on phase-field fracture},
    pdfauthor={Wei Wang, Meenu Krishnan U., Somdatta Goswami}
]{hyperref}
\usepackage[nameinlink,capitalize]{cleveref} %

\title{\scalebox{1.4}{\textbf{Axion}}: A differentiable, GPU-native finite element framework, demonstrated on phase-field fracture}

\author{
Wei Wang$^{1,2,3}$\thanks{These authors contributed equally to this work.}
\and
Meenu Krishnan U.$^{1,2,3}$\footnotemark[1]
\and
Somdatta Goswami$^{1,2,3}$\thanks{Corresponding author: \texttt{somdatta@jhu.edu}}
} 

\date{\small
$^1$Department of Civil and Systems Engineering, Johns Hopkins University\\
$^2$Hopkins Extreme Materials Institute, Johns Hopkins University\\
$^3$Data Science and AI Institute, Johns Hopkins University
}
\begin{document}
\maketitle

\begin{abstract}
\noindent
We introduce Axion (Autodifferentiable, XLA-compiled, Implicit, On-device Numerics), a differentiable, GPU-native finite element framework written in JAX. The element kernels, the Newton and Krylov iterations, the bound-constrained updates, the staggered coupling of fields, and the loop over load increments are expressed as a single program that is just-in-time compiled and executed on the GPU without host intervention during the solution. Consistent tangent operators are obtained by forward-mode automatic differentiation of the element residual, and forward-mode derivatives of the discrete solution with respect to model parameters are obtained by differentiating the same program. Global operators are applied element by element without assembling a sparse matrix, octree mesh refinement is performed within fixed-size arrays, and a cell-based domain decomposition executes the same program on several GPUs. Phase-field fracture is implemented as the introductory module, because of its known complexity; It combines fine mesh resolution, bound-constrained damage, and repeated nonlinear solves over a loading history; together this combination is computationally expensive.  On two- and three-dimensional single-edge-notched tension benchmarks, a uniformly refined Axion computation completes faster than the corresponding adaptive FEniCSx computation. At identical discretization, the two-dimensional benchmark completes on one GPU 37.9 times faster than a FEniCSx implementation on one CPU. Strong and weak scaling efficiencies on four GPUs are 92\% and 96\%, respectively. Large scale complex physics examples included for demonstration comprise hydraulic fracture in a poroelastic medium ($0.38$M degrees of freedom), fracture of a multi-material premolar under contact loading ($0.69$M degrees of freedom), and a concrete gravity dam ($96$M degrees of freedom) on a single GPU. The implementation is available as open-source software.
\end{abstract}

\noindent\textbf{Keywords:} differentiable finite element framework; GPU computing; JAX; Automatic differentiation; Adaptive mesh refinement; Phase-field fracture


 
\section{Introduction}
\label{sec:intro}

Finite element analysis is increasingly used not only for forward simulation but also within inverse analysis \cite{oberai2003solution}, uncertainty quantification \cite{ghanem2003stochastic,modak2024sparse}, optimal design \cite{bendsoe1988generating}, digital twins \cite{kapteyn2021probabilistic}, and scientific machine learning \cite{lu2021learning,goswami2022physics}. These workflows require repeated evaluations of the same model, often thousands or more, together with derivatives of the solution with respect to material parameters, geometry, boundary conditions, or constitutive laws. Their cost is therefore set by the cost of repeatedly solving and differentiating a high-fidelity model rather than by the cost of a single solution. Most production finite element codes were developed for CPU architectures and rely on assembled sparse matrices and host-side solver libraries \cite{petsc-user-ref}. Consistent tangents are generated automatically in some environments \cite{UFLAlnaesEtAl2012,korelc2016automation}, but in many codes they are still derived and implemented by hand, which is laborious and error-prone for complex constitutive models. GPU execution, for faster simulation and exact calculation of gradients, is standard in machine learning, whereas its use for implicit, nonlinear finite element analysis is less mature \cite{mills2021toward}, and derivatives of the solution with respect to model inputs are, in most codes, either unavailable or obtained from separately implemented adjoint models.

Scientific machine learning (SciML), including neural operators \cite{lu2019deeponet,li2020fourier, sarkar2026learning}, foundation models \cite{herde2024poseidon}, and physics-informed learning \cite{raissi2019physics, karniadakis2021physics}, addresses this challenge by replacing repeated high-fidelity simulations with learned differentiable surrogates. Training such surrogates, however, still requires physics-based simulations \cite{hamdi2025towards}, and purely data-driven models can lose accuracy outside the range of their training data \cite{zhu2023reliable}. These limitations have motivated hybrid AI-physics approaches that couple learning directly with numerical solvers \cite{wang2025time, wang2027non, chumbipuma2026hybrid}. At scale, however, such methods remain dependent on solvers that are both efficient enough to generate training data and differentiable enough to support end-to-end learning. Several efforts move toward this goal, but each addresses only part of the problem. GPU-enabled finite element libraries such as deal.II \cite{arndt2025deal} and MFEM \cite{anderson2021mfem} provide optimized matrix-free operators \cite{kronbichler2012generic,kronbichler2019multigrid} and keep the data on the device, but the Newton iterations, constraint treatment, and load stepping are controlled from the host. Differentiable frameworks such as dolfin-adjoint \cite{mitusch2019dolfin} and JAX-FEM \cite{xue2023jaxfem} automate tangent and adjoint calculations, but they assemble sparse Jacobians and drive the Newton iterations from the host, with the linear systems delegated to external solvers such as PETSc. In these frameworks, the GPU therefore accelerates individual kernels while the solution process is controlled from the host. For phase-field fracture, matrix-free multigrid solvers have been developed for CPUs \cite{jodlbauer2020matrix},and two recent works describe GPU implementations: a matrix-free JAX framework for
quasi-static and dynamic phase-field fracture~\cite{zhang2026matrixfree}, and PhAST, a
differentiable PyTorch solver for explicit dynamic fracture in which the damage solve
is differentiated implicitly and used to identify the fracture toughness and the
regularization length~\cite{ani2026}. \Cref{tab:related} compares these frameworks with \texttt{Axion}.

\begin{table}[htbp]
\centering
\footnotesize
\setlength{\tabcolsep}{4pt}
\begin{threeparttable}
\caption{Capabilities of GPU-capable and differentiable phase-field and finite element frameworks, as stated in their primary sources (papers, documentation and source code, consulted in November 2026). Parameter derivatives are derivatives of the computed solution with respect to model parameters, obtained in reverse mode (adjoint) or in forward mode. \checkmark: available; --: not available or not reported in the sources consulted.}
\label{tab:related}
\begin{tabular*}{\textwidth}{@{\extracolsep{\fill}}lcccccc@{}}
\toprule
 & Implicit & GPU & AD & Parameter & Adaptive & Multi-GPU \\
 & quasi-static & & tangents & derivatives & refinement & \\
\midrule
JAX-FEM \cite{xue2023jaxfem} & \checkmark & \checkmark & \checkmark & reverse mode & -- & -- \\
FEAX \cite{feax} & \checkmark & \checkmark & \checkmark & reverse mode & \checkmark\tnote{a} & \checkmark\tnote{b} \\
Zhang \& Tang \cite{zhang2026matrixfree} & \checkmark & \checkmark & -- & -- & -- & -- \\
PhAST \cite{ani2026} & \checkmark & \checkmark & \checkmark & reverse mode & -- & -- \\
Ratel \cite{digioacchino2026matrixfree} & \checkmark & \checkmark & \checkmark & -- & -- & -- \\
Yu et al.\ \cite{yu2025fullprocess} & \checkmark & \checkmark & -- & -- & -- & -- \\
\texttt{Axion} (this work) & \checkmark & \checkmark & \checkmark & forward mode & \checkmark\tnote{c} & \checkmark \\
\bottomrule
\end{tabular*}
\begin{tablenotes}[flushleft]
\footnotesize
\item[a] Remeshing only between iterations of its topology-optimization tools; the mesh is fixed within a solve.
\item[b] Independent problems can be distributed over the GPUs of one machine; a single problem is not decomposed.
\item[c] Quadtree and octree refinement of box-shaped domains; not yet combined with multi-GPU execution.
\end{tablenotes}
\end{threeparttable}
\end{table}

\begin{figure}[!t]
\centering
\includegraphics{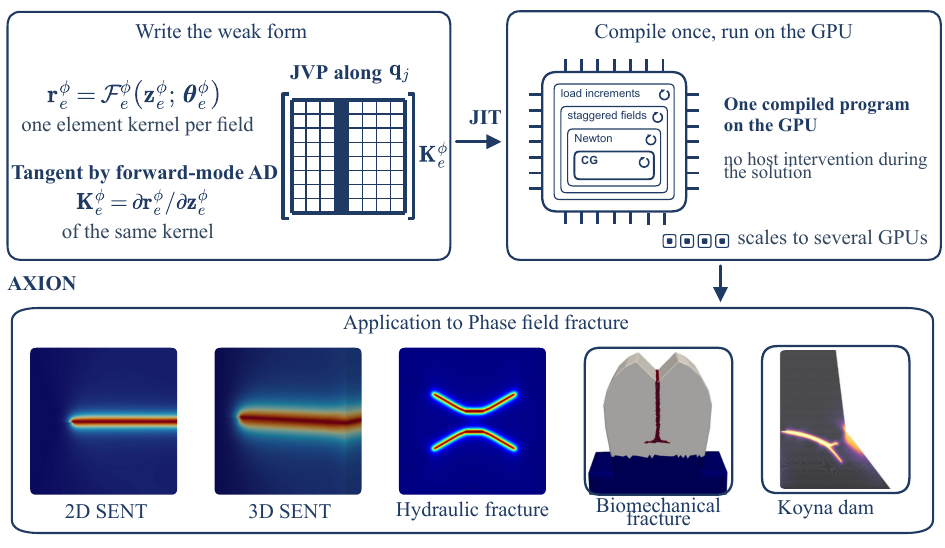}
\caption{Overview of \texttt{Axion}. Top: each field is defined by one
element kernel, whose tangent follows by forward-mode automatic differentiation
(\cref{subsec:tensorized_jit_fem}); the load, staggered, Newton and CG loops are
compiled into one GPU program (\cref{sec:compiled}) that also runs on several
GPUs (\cref{subsec:distributed_multigpu}). Bottom: the phase-field fracture
problems solved in this work (\cref{sec:examples}).}
\label{fig:overview}
\end{figure}

We present \textbf{\texttt{Axion}}
(\textbf{A}utodifferentiable, \textbf{X}LA-compiled, \textbf{I}mplicit,
\textbf{O}n-device \textbf{N}umerics), a finite element framework designed so
that finite element analysis and gradient-based learning can use the same
programming model, compiler, and hardware. \texttt{Axion} is written in
\texttt{JAX} \cite{jax2018github}, and the complete solution process,
comprising the element kernels, residual evaluation, Newton linearization, the
iterative linear solver, bound constraints, staggered coupling of fields, and
load increments, is compiled into a single program that runs on the GPU
without host intervention during the solution (\cref{sec:compiled}).
Consistent tangents are obtained by automatic differentiation of the element
residual (\cref{subsec:tensorized_jit_fem}), and forward-mode derivatives of
the discrete solution with respect to model parameters are obtained by
differentiating the same program (\cref{sec:diffmap,sec:adverify}). In this
paper, the term \emph{differentiable} refers to these two properties;
reverse-mode derivatives are discussed in \cref{sec:diffmap} but are not
provided. The objective is a solver that can be used without modification for
forward simulation, for the generation of training data, and inside
gradient-based optimization. \cref{fig:overview} summarizes how a model is written and
executed in \texttt{Axion}, and the problems solved with it in this work.

Phase-field fracture is chosen as the first application of \texttt{Axion}
because it is computationally demanding for four reasons. First, the
regularization length must be resolved by several elements along a crack path
that is not known in advance. Second, the coupled displacement and damage
equations must be solved repeatedly over the loading history. Third, the
damage field is bounded and irreversible, so that each damage update is a
constrained problem. Fourth, the tangents of the energy functionals used in
practice are laborious to derive by hand and change whenever the degradation
function or the crack geometric function is changed. As a consequence,
engineering-scale computations often involve millions of degrees of freedom,
even when adaptive or locally pre-refined meshes are used. The Koyna gravity dam (\cref{ex:koyna}) illustrates the resulting scale
separation. The cross-section is approximately $103\,\text{m}$ high and
$70\,\text{m}$ wide at the base, whereas the regularization length that sets
the width of the damage band is $\ell_o = 0.03\,\text{m}$. A uniform
discretization with $h = \ell_o/2 = 0.015\,\text{m}$ therefore requires
approximately $9.6\times10^{7}$ degrees of freedom
(\cref{fig:koyna}). Published analyses of this structure avoid this cost by
refining the mesh only in a region selected around the expected crack path
\cite{santillan2017dam}. Phase-field fracture is therefore a demanding first
test of \texttt{Axion}.

\noindent The principal contributions of this work are summarized as follows:
\begin{itemize}[leftmargin=*, nosep]
\item \texttt{Axion}, a differentiable, GPU-native finite element framework
written in \texttt{JAX}, in which the complete nonlinear solution process,
including load stepping, is compiled into a single program that executes on
the GPU (\cref{sec:framework}).
\item An implementation of the \texttt{AT2} and \texttt{PF-CZM} phase-field
models as the first physics module of the framework, in which both models
share one element definition and all tangent operators, including those of a
three-field hydro-poroelastic extension, are obtained by automatic
differentiation (\cref{sec:pff}).
\item An octree refinement scheme that operates within fixed-size arrays, so
that the mesh can be refined during a simulation with recompilation only when
a buffer capacity is exceeded, and a cell-based multi-GPU decomposition whose
operator action is identical to that of the single-device solver
(\cref{subsec:octree_adaptivity,subsec:distributed_multigpu}).
\item Verification on two- and three-dimensional fracture benchmarks,
verification of the automatically differentiated tangents and parameter
sensitivities against finite differences, strong- and weak-scaling studies on
up to four GPUs, three application examples, and the release of the
implementation as open-source software (\cref{sec:examples}).
\end{itemize}

The remainder of this paper is organized as follows. Section~\ref{sec:framework} introduces the \texttt{Axion} framework, including its differentiable finite element architecture and the GPU-native implementation, and Section~\ref{sec:pff} formulates phase-field fracture and shows how it is expressed in \texttt{Axion}. Section~\ref{sec:benchmarks} verifies the framework against canonical fracture benchmarks and evaluates its accuracy, computational performance, and scalability. Section~\ref{sec:largescale} demonstrates the capability of \texttt{Axion} on engineering-scale problems from industrial, geophysical, and biomechanical applications. Finally, Section~\ref{sec:conclusions} concludes the paper and outlines future research directions.
%
\section{The \texttt{Axion} framework}\label{sec:framework}

This section presents \texttt{Axion} independently of any particular physical
system. The phase-field fracture problem used to evaluate the framework is
formulated in \cref{sec:pff}, where it is also shown how that problem is
expressed in this framework.

\subsection{Overview and design principles}\label{sec:overview}

\begin{figure}[!htb]
\centering
\includegraphics[width=\textwidth]{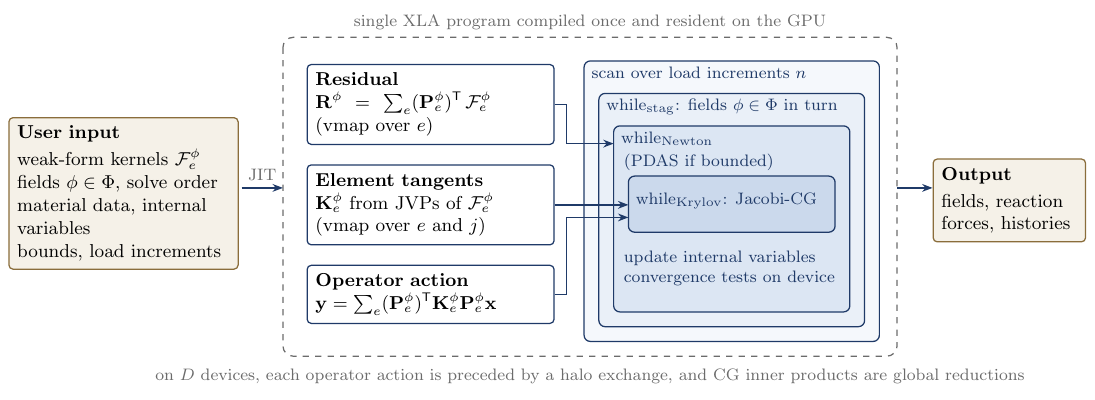}
\caption{Architecture of \texttt{Axion}. The user supplies weak-form kernels
$\mathcal{F}^{\phi}_{e}$ for each field $\phi\in\Phi$ together with material
data and loading. The framework generates the residuals, the element tangents
(by forward-mode automatic differentiation), and the operator action, and
compiles them, together with the nested solution loops, into a single XLA
program that runs on one or $D$ GPUs.}
\label{fig:architecture}
\end{figure}

\texttt{Axion} is implemented in \texttt{JAX} \cite{jax2018github}, which
provides automatic differentiation (AD), vectorized mapping (\texttt{vmap}),
and just-in-time (JIT) compilation through the XLA compiler within a
NumPy-style array language. A coupled problem is represented as a set of
subproblems, one for each field equation, indexed by $\phi\in\Phi$; for
phase-field fracture, $\Phi=\{u,d\}$ denotes the displacement and damage
equations. Each subproblem has a vector of nodal unknowns $\mathbf{z}^{\phi}$
and is defined by an element-level weak-form kernel $\mathcal{F}^{\phi}_{e}$
that returns the element residual. The user provides these kernels, the
material data, the internal variables, the boundary constraints, the order in
which the subproblems are solved, and the sequence of load increments.
\texttt{Axion} generates the residual evaluation, the tangent construction,
the linear solution, the treatment of constraints, and the solution control.
The framework rests on three design principles (\cref{fig:architecture}).
\begin{enumerate}[leftmargin=*, nosep]
\item \emph{One element definition per field.}
The element residual and its consistent tangent are obtained from the same
kernel $\mathcal{F}^{\phi}_{e}$, the tangent by automatic differentiation
(\cref{subsec:tensorized_jit_fem}). Global operators are applied element by
element from the stored element tangents, without assembling a sparse matrix
(\cref{sec:matfree}).
\item \emph{Fixed array shapes.}
All array shapes are fixed at compilation. Quantities whose size may change
during a simulation, such as the active cells of a refined mesh, are stored in
fixed-capacity buffers with masks (\cref{subsec:octree_adaptivity}), and bound
constraints are enforced by projection (\cref{sec:pdas}), so that the logical
problem can change without a change in the compiled array shapes.
\item \emph{Compiled control flow.}
The load increments and the staggered, Newton, and Krylov iterations are
expressed with \texttt{JAX} control-flow primitives, so that the complete
solution process is a single XLA program that remains on the GPU; the host is
involved only before and after the simulation (\cref{sec:compiled}).
\end{enumerate}

\subsection{Element kernels and automatically generated tangents}
\label{subsec:tensorized_jit_fem}
The first step in this execution model is to express the finite element
residual as a vectorized element operation. Let $N_{\mathrm{cell}}$ denote the number of elements and $n^{\phi}_{e}$ the
number of unknowns of field $\phi$ on one element. For an element $e$, the
local unknown vector
$\mathbf{z}^{\phi}_{e}=\mathbf{P}^{\phi}_{e}\mathbf{z}^{\phi}\in\mathbb{R}^{n_e^{\phi}}$
is extracted from the global vector by the Boolean gather operator
$\mathbf{P}^{\phi}_{e}$. The kernel $\mathcal{F}^{\phi}_{e}$ evaluates the
isoparametric interpolation, the constitutive response at the quadrature
points, and the weighted quadrature sum, and returns the element residual
$\mathbf{r}^{\phi}_{e}$, from which the global residual $\mathbf{R}^{\phi}$
follows by standard finite element assembly,
\begin{equation}
\mathbf{r}^{\phi}_{e}
=\mathcal{F}^{\phi}_{e}\left(\mathbf{P}^{\phi}_{e}\mathbf{z}^{\phi};\boldsymbol{\theta}^{\phi}_{e}\right),
\qquad
\mathbf{R}^{\phi}\left(\mathbf{z}^{\phi};\boldsymbol{\theta}^{\phi}\right)
=\sum_{e=1}^{N_{\mathrm{cell}}}\left(\mathbf{P}^{\phi}_{e}\right)^{T}\mathbf{r}^{\phi}_{e},
\label{eq:global_residual_assembly}
\end{equation}
where $\boldsymbol{\theta}^{\phi}_{e}$ contains quadrature-point geometry,
material parameters, internal variables, and the fields of the other
subproblems, which remain fixed while field $\phi$ is solved, and
$\boldsymbol{\theta}^{\phi}=\{\boldsymbol{\theta}^{\phi}_{e}\}_{e=1}^{N_{\mathrm{cell}}}$
collects these data over the mesh.

In conventional implementations, the element evaluations of \cref{eq:global_residual_assembly} are performed in a loop over the elements on the host. In
\texttt{Axion}, all element-local quantities are stored in arrays whose first dimension indexes the elements,
so the element loop becomes a single vectorized array operation that can be
compiled by JAX. Newton linearization requires the element tangent
$\mathbf{K}^{\phi}_{e}=\partial\mathbf{r}^{\phi}_{e}/\partial\mathbf{z}^{\phi}_{e}$.
Rather than deriving it by hand, \texttt{Axion} computes its
columns as forward-mode Jacobian-vector products (JVPs)
\cite{griewank2008evaluating} of the \emph{same} kernel
$\mathcal{F}^{\phi}_{e}$ along the canonical basis vectors
$\{\mathbf{q}_j\}_{j=1}^{n_e^{\phi}}$ of $\mathbb{R}^{n_e^{\phi}}$. Residual
and tangent are thus produced by vectorized maps over the element axis, and
the tangent additionally over the basis vectors,
\begin{align}
&\mathbf{R}^{\phi}_{\mathrm{cell}}
=\operatorname{vmap}_{e}\left[\mathcal{F}^{\phi}_{e}\left(\mathbf{z}^{\phi}_{e};\boldsymbol{\theta}^{\phi}_{e}\right)\right],
\label{eq:tensorized_residual_vmap}\\
&\left\{\mathbf{K}^{\phi}_{e}\right\}_{e=1}^{N_{\mathrm{cell}}}
=\operatorname{vmap}_{e}\left[\operatorname{vmap}_{j}\left(\operatorname{JVP}\left[\mathcal{F}^{\phi}_{e},\mathbf{z}^{\phi}_{e};\mathbf{q}_j\right]\right)\right],
\label{eq:vectorized_tangent_construction}
\end{align}
where $\mathbf{R}^{\phi}_{\mathrm{cell}}$ is the array of all element
residuals before assembly, and
$\operatorname{JVP}[\mathcal{F},\mathbf{z};\mathbf{q}]$ denotes the
directional derivative of $\mathcal{F}$ at $\mathbf{z}$ in the direction
$\mathbf{q}$. The nonlinear residual and its consistent tangent therefore originate from a single element definition: any change to the weak
form, whether a different constitutive law, degradation function, or coupling
term, propagates to the tangent automatically. The tangent is the derivative of the implemented residual to round-off, so that Newton's method retains quadratic convergence where the residual is smooth (\cref{sec:adverify}). For trilinear hexahedral (HEX8)
elements with $2\times2\times2$ Gauss quadrature, for example, a vector-valued
field has $n_e^{\phi}=24$ and a scalar field $n_e^{\phi}=8$ local unknowns. The present implementation provides linear triangles (TRI3) and tetrahedra (TET4), bilinear quadrilaterals (QUAD4), and trilinear hexahedra (HEX8), and the element family used in each example is stated in \cref{sec:examples}.

\subsection{Element-by-element operator evaluation and linear solvers}
\label{sec:matfree}

The global tangent,
$\mathbf{K}^{\phi}=\sum_e(\mathbf{P}^{\phi}_{e})^{T}\mathbf{K}^{\phi}_{e}\mathbf{P}^{\phi}_{e}$,
is represented implicitly through its element
contributions rather than explicitly as a sparse matrix. This avoids the construction of a sparsity pattern and the assembly of a
global matrix, and it keeps all array shapes independent of the mesh
connectivity, which the adaptive scheme of \cref{subsec:octree_adaptivity}
requires. It does not reduce storage: for trilinear hexahedra, the $576$
values stored per element exceed the approximately $243$ non-zero entries per
node of the assembled displacement tangent, and a hexahedral mesh has about
one element per node. In contrast to matrix-free methods based on sum
factorization \cite{kronbichler2012generic,kronbichler2019multigrid}, the
element tangents are computed once per Newton iteration and stored, an
approach known as element-by-element evaluation \cite{hughes1983element}. Each operator application consists of three steps: gathering the entries of
$\mathbf{x}$ associated with each element, performing the local
matrix-vector products $\mathbf{K}^{\phi}_{e}\mathbf{x}_e$ \cite{hughes1983element},
\begin{equation}
\mathbf{y}=\mathbf{K}^{\phi}\mathbf{x}
=\sum_{e=1}^{N_{\mathrm{cell}}}\left(\mathbf{P}^{\phi}_{e}\right)^{T}\mathbf{K}^{\phi}_{e}\mathbf{P}^{\phi}_{e}\mathbf{x},
\label{eq:matrix_free_operator_action}
\end{equation}
and
scatter-adding the resulting contributions to the global vector. These
operations are performed as batched array operations in JAX and can
therefore be compiled and executed on the GPU without constructing a global
sparse data structure. On a fixed mesh, the local-to-global index maps remain
unchanged between Newton iterations, so only the element tangent values need
to be updated. The storage required is $(n_e^{\phi})^2$ values per element,
corresponding to $576$ double-precision numbers ($4.6$\;kB) for the
displacement field of a HEX8 element.

The linearized systems are solved by the conjugate gradient (CG) method
\cite{hestenes1952methods} with a Jacobi preconditioner, whose diagonal is
obtained by scatter-adding the diagonals of the element tangents. CG requires
the system matrix to be symmetric positive definite; this holds for the
elasticity and pressure subproblems considered in this work, and for the
damage subproblems it is discussed in \cref{sec:pff-in-axion}. Dirichlet constraints are imposed by symmetric elimination, which preserves
symmetry without modifying the stored tangent values. Solver tolerances are
chosen per problem; the values quoted here and in \cref{sec:compiled} are those
of the three-dimensional benchmark of \cref{ex:sent3d}, which we take as the
representative configuration. There, the CG iteration is terminated when
$\|\mathbf{r}\|_2/\|\mathbf{b}\|_2<10^{-6}$ and the Newton iteration when
$\|\mathbf{R}^{\phi}\|_2<10^{-8}$ or after $20$ iterations. Across the remaining
examples the CG tolerance ranges from $10^{-8}$ to $10^{-4}$ and the Newton cap
from $20$ to $30$ iterations.

\subsection{Newton iterations and bound constraints}\label{sec:pdas}

Each nonlinear subproblem is advanced by Newton iterations, indexed by $j$,
with $\boldsymbol{\theta}^{\phi}$ held fixed and omitted from the arguments,
\begin{equation}
\mathbf{K}^{\phi}\left(\mathbf{z}^{\phi}_{j}\right)\Delta\mathbf{z}^{\phi}_{j}
=-\mathbf{R}^{\phi}\left(\mathbf{z}^{\phi}_{j}\right),
\qquad
\mathbf{z}^{\phi}_{j+1}=\mathbf{z}^{\phi}_{j}+\Delta\mathbf{z}^{\phi}_{j},
\label{eq:newton_update}
\end{equation}
where each linear system is solved by the preconditioned CG method of
\cref{sec:matfree}. A subproblem may be subject to box constraints
$\underline{\mathbf{z}}\le\mathbf{z}^{\phi}\le\overline{\mathbf{z}}$, as the
damage field is in phase-field fracture. If the subproblem is linear and its
unconstrained solution satisfies the bounds up to discretization error, a
single solve of \cref{eq:newton_update} followed by a projection onto the
admissible set is used. Otherwise, in particular when the subproblem is
non-convex and its tangent indefinite, it is solved by a projected Jacobi
iteration \cite{bertsekas1989parallel}, indexed by $m$,
\begin{equation}
\mathbf{z}^{\phi}_{m+1}
=\Pi\!\left(\mathbf{z}^{\phi}_{m}-\mathbf{D}^{-1}\mathbf{R}^{\phi}\big(\mathbf{z}^{\phi}_{m}\big)\right),
\qquad
D_{ii}=\big|K^{\phi}_{ii}\big|,
\label{eq:projected_jacobi}
\end{equation}
where $\Pi$ is the componentwise projection onto
$[\underline{\mathbf{z}},\overline{\mathbf{z}}]$ and the diagonal of
$\mathbf{K}^{\phi}$ is assembled by scatter-adding the diagonals of the element
tangents of \cref{eq:vectorized_tangent_construction}. The iteration requires
no linear solve, and the absolute value makes the scaling positive regardless
of the sign of the curvature, so no definiteness of the tangent is needed. A
fixed point of \cref{eq:projected_jacobi} satisfies the first-order optimality
conditions of the bound-constrained problem: $R^{\phi}_i=0$ where $z^{\phi}_i$
lies strictly between its bounds, $R^{\phi}_i\ge0$ at the lower bound, and
$R^{\phi}_i\le0$ at the upper bound. The unknowns held at a bound are encoded by the projection itself, so all array
shapes are invariant, and the iteration is a compiled while-loop, terminated
when the relative change of $\mathbf{z}^{\phi}$ falls below $10^{-7}$ or after a
prescribed maximum number of sweeps. The premolar computation of
\cref{ex:molar} is the only example in this work driven by
\cref{eq:projected_jacobi}, and it meets this criterion only before the crack
propagates. Over the first $40$ load increments the relative change at
termination lies between $3\times10^{-14}$ and $1\times10^{-7}$. From increment $41$
($\bar\delta=0.164\,\text{mm}$) onwards, as the crack advances through the
enamel, the iteration stalls: every subsequent increment reaches the sweep cap
of $2000$ with the relative change held between $5\times10^{-3}$ and
$7\times10^{-2}$, and $31\%$ of all damage solves in the computation terminate
at the cap rather than at the tolerance. The staggered iteration is limited by
its own cap of $K_{\max}=60$ from increment $33$ ($\bar\delta=0.132\,\text{mm}$)
onwards. Because a small change between successive Jacobi iterates does not by
itself imply a small residual, this behaviour is a limitation of the projected
Jacobi iteration for the \texttt{PF-CZM} degradation function, whose tangent is
indefinite in the softening regime; a projected residual, which would vanish
exactly at points satisfying the optimality conditions above, is not evaluated
in the present implementation.

\subsection{The solution process as a single compiled program}
\label{sec:compiled}

Coupled subproblems are solved by staggered (alternating) iteration: within
each load increment, the subproblems $\phi\in\Phi$ are solved in the
prescribed order, each with its data $\boldsymbol{\theta}^{\phi}$ held fixed,
and the sweep is repeated until the fields no longer change
(\cref{alg:staggered_phase_field}). The data $\boldsymbol{\theta}^{\phi}$ are
formed from the most recent values of the other fields, namely the $(k+1)$
iterates of the subproblems already solved in the current sweep and the $(k)$
iterates of the remaining ones, so that the scheme is a block Gauss--Seidel iteration, as in the alternate
minimization of variational phase-field models \cite{farrell2017linear}. In the representative configuration of \cref{ex:sent3d},
$\varepsilon_{\mathrm{stag}}=10^{-3}$ and $K_{\max}=8$; the other examples use
problem-specific values, with $\varepsilon_{\mathrm{stag}}$ between $10^{-5}$
and $10^{-3}$ and $K_{\max}$ between $8$ and $1000$.
 The entire workflow is expressed with JAX control-flow primitives: the load
increments form a compiled $\operatorname{scan}$, while the staggered, Newton,
and Krylov iterations are compiled while-loops,
\begin{equation}
\operatorname{scan}_{n}\left[\operatorname{while}_{\mathrm{stag}}\left[\operatorname{while}_{\mathrm{Newton}}\left[\operatorname{while}_{\mathrm{Krylov}}\right]\right]\right].
\label{eq:compiled_solver_hierarchy}
\end{equation}
Static compilation does not fix the number of solver iterations: only the
shapes, data types, and tree structures of the variables carried from one
iteration to the next are invariant, whereas all termination criteria are
evaluated on the device at run time. On a fixed mesh every kernel is therefore compiled exactly once, the complete
finite element state stays in device memory across all load increments, and no
host--device synchronization occurs during the nonlinear solution. This is the
mode in which the benchmarks of \cref{sec:benchmarks} are run. 
 This property distinguishes \texttt{Axion} from frameworks in which the data
reside on the GPU but the solution process is controlled from the host
(\cref{sec:intro}). Its contribution to the run times reported in
\cref{sec:examples} is not isolated in this work.
Compilation takes $8.2\,\text{s}$ for the two-dimensional benchmark and
$2.2\,\text{min}$ for the three-dimensional benchmark, in both cases dominated
by the \texttt{XLA} compile rather than by tracing, and the times reported in
\cref{tab:cost} include compilation, which accounts for $3.0\%$ and $3.8\%$ of
them respectively. Because the load loop is a $\operatorname{scan}$, which
lowers to a single \texttt{XLA} while-loop, this cost is independent of the
number of load increments and is therefore amortized over the whole loading
history.

\begin{algorithm}[!htb]
\caption{Staggered solution of coupled subproblems within one compiled program}
\label{alg:staggered_phase_field}
\begin{algorithmic}[1]
\Require load increments $n=1,\ldots,N$; ordered fields $\Phi$; tolerance
$\varepsilon_{\mathrm{stag}}$; maximum number of staggered iterations
$K_{\max}$; small constant $\epsilon>0$
\For{$n=1,\ldots,N$} \Comment{compiled $\operatorname{scan}$}
    \State apply the boundary data of increment $n$; initialize all fields from increment $n-1$; $k\gets 0$, $\Delta_{\mathrm{stag}}\gets\infty$
    \While{$\Delta_{\mathrm{stag}}>\varepsilon_{\mathrm{stag}}$ \textbf{and} $k<K_{\max}$} \Comment{compiled while-loop}
            \For{$\phi\in\Phi$}
        \State form $\boldsymbol{\theta}^{\phi}$ from the latest iterates of the other fields and the current internal variables
       \State solve $\mathbf{R}^{\phi}\big(\mathbf{z}^{\phi,(k+1)};\boldsymbol{\theta}^{\phi}\big)=\mathbf{0}$ by \cref{eq:newton_update}, or by \cref{eq:projected_jacobi} if bounded and non-convex
        \State update the internal variables that depend on $\mathbf{z}^{\phi,(k+1)}$
    \EndFor
        \State $\Delta_{\mathrm{stag}}\gets\max_{\phi}\big\|\mathbf{z}^{\phi,(k+1)}-\mathbf{z}^{\phi,(k)}\big\|_2\big/\big(\big\|\mathbf{z}^{\phi,(k+1)}\big\|_2+\epsilon\big)$;\quad $k\gets k+1$
    \EndWhile
    \State accept the increment; store reaction forces and other output quantities
\EndFor
\end{algorithmic}
\end{algorithm}

\subsection{Adaptive refinement with static array shapes}
\label{subsec:octree_adaptivity}

Solutions with localized features, such as boundary layers, shear bands, or
cracks, are resolved economically on hierarchically refined quadtree or octree
meshes \cite{burstedde2011p4est,bangerth2007dealii,heister2015primal}. In a
JIT-compiled setting, however, every change in array shape would trigger
recompilation of all GPU kernels. The adaptivity in \texttt{Axion} is
therefore designed around one requirement: \emph{the logical mesh may evolve
while the compiled array shapes remain static}. The present implementation
applies to box-shaped domains discretized by quadrilaterals or hexahedra, and
supports refinement but not coarsening.

\paragraph{Hierarchy and device-side node identity}
The domain $(0,L_x)\times(0,L_y)\times(0,L_z)$ is discretized by an initial
Cartesian lattice of $N_{x,0}\times N_{y,0}\times N_{z,0}$ hexahedra and
refined up to a maximum level $L_{\max}$; a cell at refinement level $L_e$ has
dimensions $h_x^{(e)}=L_x/(N_{x,0}2^{L_e})$, and similarly in $y$ and $z$. In
two dimensions the construction reduces to a quadtree with $1\to4$ refinement.
Node identities are defined once and for all on the finest possible lattice:
with $N_x=N_{x,0}2^{L_{\max}}$ and $N_y$, $N_z$ defined analogously, the node
at integer lattice coordinates $(I,J,K)$ carries the key
\begin{equation}
\kappa(I,J,K)=\left[I(N_y+1)+J\right](N_z+1)+K .
\label{eq:lattice_key}
\end{equation}
Duplicate nodes generated simultaneously by neighboring refined cells are
therefore merged by integer comparison alone: node creation, neighbor
queries, and the balance sweeps below need no geometric tolerance, no hash
table, and no host-side search, and are executed as sorted-array operations on
the device.

\paragraph{Marking, balance, and constraints}
Refinement is driven by a user-supplied cell-wise indicator $\eta_e\ge 0$: a
cell is marked when $\eta_e$ exceeds a threshold $\eta_{\mathrm{th}}$,
provided it has reached neither the target size $h_{\min}$ nor the maximum
level $L_{\max}$. Each refinement pass is followed by a 2:1 balance sweep,
$|L_e-L_{e'}|\le 1$ for neighboring active cells $e$ and $e'$, which ensures
that every hanging node lies on an edge or face of a coarser neighbor and
admits the standard interpolation constraints \cite{bangerth2007dealii}: an
edge hanging node takes the mean of its two master nodes, and a face-center
node the mean of its four. Eliminating the constrained and Dirichlet unknowns
writes the field as $\mathbf{z}=\mathbf{T}\widehat{\mathbf{z}}+\mathbf{c}$,
where $\widehat{\mathbf{z}}$ collects the free unknowns, $\mathbf{T}$ is the
constraint matrix, and $\mathbf{c}$ contains the prescribed values. The
linearized system $\mathbf{K}\mathbf{z}=\mathbf{b}$ then becomes
$\mathbf{T}^{T}\mathbf{K}\mathbf{T}\widehat{\mathbf{z}}=\mathbf{T}^{T}(\mathbf{b}-\mathbf{K}\mathbf{c})$,
a congruence transformation that preserves symmetry and positive
definiteness. Because $\mathbf{T}$ and $\mathbf{T}^{T}$ are sparse
scatter-add maps, they are incorporated directly into
\cref{eq:matrix_free_operator_action}, and the preconditioned CG solver runs
unchanged on nonconforming meshes.

\paragraph{Fixed-capacity arenas and bucketed execution}
The mesh is stored in padded buffers of fixed capacity, called \emph{arenas},
which hold the node keys, coordinates, connectivities, refinement levels,
validity flags, and cell-wise internal variables; inactive and padded entries
are masked out of every element evaluation. Refinement only appends entries,
with an occasional compaction that preserves the buffer size, so the
adaptation cycle runs entirely inside compiled control flow. Values at new
nodes follow from trilinear interpolation within the parent cell, which is
exact because the refined space contains the coarse one, and when a cell is refined, the quadrature-point internal variables of its
children are copied from the nearest quadrature point of the parent, so that the history field
is not reduced by refinement. A run therefore
proceeds in \emph{capacity buckets}: many load increments share one compiled
program, and the kernels are recompiled only when the number of active
entries exceeds a prescribed fraction $\tau$ of the capacity $\mathcal{C}$, at
which point the capacity is enlarged by a factor $\gamma$. The cost of
compilation is thus amortized over many increments while the mesh grows with
the solution (\cref{alg:adaptive_bucket}, in which $\mathcal{S}_{n-1}$ denotes
the complete solution state at the end of increment $n-1$). In the adaptive computation of \cref{ex:sent} the arenas were allocated at twice
the size of the initial mesh, which exceeds the peak requirement of the run:
occupancy reaches $77\%$ of the cell capacity and $78\%$ of the node capacity
against a growth threshold of $\tau=0.9$, so no capacity growth is triggered and
the kernels are compiled exactly once, in $9.0\,\text{s}$. The reported
$2.41\,\text{min}$ therefore contains no recompilation. Growth is exercised only
when the arenas are deliberately started below the peak, for which
$\gamma=1.1$ is used.

\begin{algorithm}[!htb]
\caption{Adaptive load increment with static-shape capacity buckets}
\label{alg:adaptive_bucket}
\begin{algorithmic}[1]
\Require arena capacity $\mathcal{C}$, growth factor $\gamma>1$, fill fraction $\tau\in(0,1)$, refinement threshold $\eta_{\mathrm{th}}$
\State allocate arenas of capacity $\mathcal{C}$; compile all kernels for these shapes
\For{$n=1,\ldots,N$}
    \State store the state $\mathcal{S}_{n-1}$
    \State solve increment $n$ on the current mesh (\cref{alg:staggered_phase_field})
    \For{$s=1,\ldots,S$} \Comment{fixed number of sweeps, $S=8$ in this work}
        \State mark cells with $\eta_e>\eta_{\mathrm{th}}$ whose size exceeds $h_{\min}$
        \State refine the marked cells; restore the 2:1 balance; rebuild $(\mathbf{T},\mathbf{c})$; interpolate the fields
        \State solve increment $n$ again on the current mesh \Comment{also when no cell was marked}
    \EndFor
    \If{the peak number of active entries exceeded $\tau\,\mathcal{C}$}
        \State $\mathcal{C}\gets\lceil\gamma\,\mathcal{C}\rceil$; compact, reallocate, recompile; restore $\mathcal{S}_{n-1}$ and repeat increment $n$ \Comment{new bucket}
    \EndIf
\EndFor
\end{algorithmic}
\end{algorithm}

\subsection{Multi-GPU execution by cell decomposition}
\label{subsec:distributed_multigpu}

For computations on $D$ devices, \texttt{Axion} adopts a cell-based domain
decomposition \cite{smith1996domain}: the $N_{\mathrm{cell}}$ active cells are
split into disjoint subdomains $\mathcal{T}_h^{(p)}$, $p=1,\ldots,D$, with
$|\mathcal{T}_h^{(p)}|\approx N_{\mathrm{cell}}/D$, computed either by
multilevel graph partitioning with \textsc{METIS} \cite{karypis1998fast}, by a
Morton space-filling-curve ordering \cite{bader2013space}, or by a division
into coordinate slabs for box-shaped domains. Device $p$ stores the unknowns
it owns, with index set $\mathcal{I}_{p}^{\phi,\mathrm{own}}$, together with a
one-layer ghost copy of the interface unknowns it touches but does not own,
$\mathcal{I}_{p}^{\phi,\mathrm{gh}}$, so that its local index set is
$\mathcal{I}_{p}^{\phi,\mathrm{loc}}=\mathcal{I}_{p}^{\phi,\mathrm{own}}\cup\mathcal{I}_{p}^{\phi,\mathrm{gh}}$.

\paragraph{Distributed operator action}
Each device evaluates element quantities only for its own cells. For an
operator application, a halo exchange first forms the local vector
$\mathbf{x}_{p}^{\phi,\mathrm{loc}}$ from the owned values and the ghost
values received from neighboring devices; each device then computes its
unassembled contribution, and the contributions to interface rows are
accumulated on their owning devices,
\begin{equation}
\widetilde{\mathbf{y}}_{p}^{\phi}
=\sum_{\varOmega_e\in\mathcal{T}_h^{(p)}}\left(\mathbf{P}_{e,p}^{\phi}\right)^{T}\mathbf{K}_{e}^{\phi}\mathbf{P}_{e,p}^{\phi}\,\mathbf{x}_{p}^{\phi,\mathrm{loc}},
\qquad
\mathbf{y}_{p}^{\phi,\mathrm{own}}
=\sum_{q=1}^{D}\mathbf{C}_{p\leftarrow q}^{\phi}\,\widetilde{\mathbf{y}}_{q}^{\phi},
\label{eq:distributed_matvec_assembly}
\end{equation}
where $\mathbf{P}_{e,p}^{\phi}$ extracts the unknowns of element $e$ from the
local vector of device $p$, and $\mathbf{C}_{p\leftarrow q}^{\phi}$ transfers
the entries computed on device $q$ whose rows are owned by device $p$ and adds
them to $p$'s result. The residual of \cref{eq:global_residual_assembly} is
distributed by the same pattern. The distributed operator action is therefore
mathematically identical to \cref{eq:matrix_free_operator_action}, merely
decomposed into local element evaluation, halo exchange, and distributed
accumulation. The inner products and norms required by the Krylov solver are
reduced over all devices in the same way, so that the staggered, Newton, and
Krylov convergence histories coincide with the single-device ones. The
computation is expressed over a one-dimensional logical mesh of devices using
the sharding facilities of JAX, which are lowered to the GSPMD partitioner of
XLA \cite{xu2021gspmd}; multi-node runs use the JAX distributed runtime and the
vendor collective-communication libraries. Every device executes the same
compiled program on its own subdomain (single program, multiple data), which
requires identical padded array shapes on all devices: the owned-cell and
ghost counts are padded to their maxima over the devices, and element
evaluation proceeds in fixed-size batches of $B$ cells, which bounds the
memory needed for assembly regardless of the subdomain size.

\paragraph{Memory and runtime models}
Let $N_{\mathrm{dof}}$ be the total number of unknowns and $N_{\Gamma}^{(p)}$
the number of interface unknowns of device $p$. The memory footprint and the
time per iteration on device $p$ are approximately
\begin{align}
&M_p\approx M_{\mathrm{base}}+c_{\mathrm{dof}}\frac{N_{\mathrm{dof}}}{D}+c_{\mathrm{cell}}\frac{N_{\mathrm{cell}}}{D}+c_{\mathrm{ghost}}N_{\Gamma}^{(p)}+c_{\mathrm{batch}}B,
\label{eq:distributed_memory_model}\\
&T_p\approx T_{\mathrm{elem}}\!\left(\frac{N_{\mathrm{cell}}}{D}\right)+T_{\mathrm{halo}}\!\left(N_{\Gamma}\right)+T_{\mathrm{red}}\!\left(D\right),
\label{eq:distributed_time_model}
\end{align}
where $M_{\mathrm{base}}$ is the device-independent memory of the compiled
program, the constants $c_{(\cdot)}$ are the storage per unknown, per cell,
per ghost unknown, and per batched cell, and $T_{\mathrm{elem}}$,
$T_{\mathrm{halo}}$, and $T_{\mathrm{red}}$ are the costs of local element
work, interface exchange, and global reductions, respectively. Field and cell
data decrease as $1/D$, whereas for compact three-dimensional partitions the
interface obeys surface-to-volume scaling,
$N_{\Gamma}^{(p)}=\mathcal{O}[(N_{\mathrm{cell}}/D)^{2/3}]$, and hence
$N_{\Gamma}=\sum_p N_{\Gamma}^{(p)}=\mathcal{O}(D^{1/3}N_{\mathrm{cell}}^{2/3})$,
which remains small relative to $N_{\mathrm{dof}}$. Near-ideal scaling is therefore expected as long as local element and Krylov
work dominate the halo and reduction terms. These estimates are qualitative;
their terms are not fitted to measurements in this work.

\subsection{Differentiability of the solution map}\label{sec:diffmap}

Because the residuals, tangents, and solvers are written in \texttt{JAX}, the
map from input parameters $\mathbf{p}$, such as material constants, to the
discrete solution $\mathbf{z}(\mathbf{p})$ is itself a \texttt{JAX} program,
and forward-mode derivatives $\partial\mathbf{z}/\partial\mathbf{p}$ are
obtained by applying \texttt{jax.jvp} to the complete solution process of
\cref{eq:compiled_solver_hierarchy}. The cost of one such derivative is that of
carrying a tangent through every operation of the primal computation, which is
a small constant multiple of the primal cost for each parameter
\cite{griewank2008evaluating}. Differentiating the Newton and CG iterations
directly yields the derivatives of the iterates, which converge to the
derivative of the solution as the iterations converge
\cite{griewank2008evaluating}; the accuracy of the computed derivatives
therefore depends on the solver tolerances, and it is assessed in
\cref{sec:adverify}.

Internal variables defined by a maximum over the loading history, such as the
history field of \cref{sec:formulation}, the projection onto the bounds, and
changes of the active set make the solution map piecewise smooth. Derivatives
exist along a loading path except at the instants at which the history
variable switches between maximizing states or a bound becomes active or
inactive. Under monotonic loading before the peak load, the history field
equals the current tensile energy at every quadrature point and, for the
\texttt{AT2} model, no damage unknown reaches the upper bound, so that the
derivatives are well defined in that range; both conditions are checked during
the computation. Reverse-mode derivatives, which are preferable when the number
of parameters is large, are not provided in this work, because \texttt{JAX}
does not support reverse-mode differentiation of its while-loop primitive. They
can be obtained by implicit differentiation of each converged increment
\cite{blondel2022}, which requires one adjoint solve per increment, and this is
left for future work on optimization, control, and inverse analysis.

\section{Demonstration of Phase-field fracture in \texttt{Axion}}
\label{sec:pff}

This section formulates the phase-field fracture problem used to evaluate the
framework, shows how it is expressed in \texttt{Axion}, and extends it to
hydraulic fracture as an example of adding a physical field.

\subsection{Variational phase-field formulation}\label{sec:formulation}

The phase-field approach represents fracture through a smooth scalar field
that regularizes the sharp crack surface, thereby avoiding explicit crack
tracking. Consider a linear elastic body
$\varOmega\subset\mathbb{R}^{\delta}$, $\delta\in\{2,3\}$, with boundary
$\partial\varOmega=\partial\varOmega_{D}\cup\partial\varOmega_{N}$ and outward
unit normal $\boldsymbol{n}$, on which prescribed displacements
$\boldsymbol{u}^{*}$ are imposed on $\partial\varOmega_{D}$ and prescribed
tractions $\boldsymbol{t}^{*}$ on $\partial\varOmega_{N}$, and which is
subject to a body force $\boldsymbol{b}^{*}$. Two fields are defined at each
point $\boldsymbol{x}\in\varOmega$: the displacement
$\boldsymbol{u}(\boldsymbol{x})$, with small-strain tensor
$\boldsymbol{\varepsilon}=\tfrac12(\nabla\boldsymbol{u}+\nabla\boldsymbol{u}^{T})$,
and the phase field $d(\boldsymbol{x})\in[0,1]$, with $d=0$ for intact and
$d=1$ for fully broken material.

Following the variational theory of brittle fracture
\cite{francfort1998revisiting,Bourdin2000}, the fields are obtained by
minimizing the total energy
\begin{equation}\label{eq:total-energy}
\mathcal{E}(\boldsymbol{u},d)
=\int_{\varOmega}\psi_{s}(\boldsymbol{\varepsilon},d)\,\mathrm{d}V
+\int_{\varOmega}\frac{G_{\mathrm{f}}}{c_{\alpha}}\left[\frac{\alpha(d)}{\ell_o}+\ell_o\,\nabla d\cdot\nabla d\right]\mathrm{d}V
-\int_{\varOmega}\boldsymbol{b}^{*}\cdot\boldsymbol{u}\,\mathrm{d}V
-\int_{\partial\varOmega_{N}}\boldsymbol{t}^{*}\cdot\boldsymbol{u}\,\mathrm{d}A,
\end{equation}
which comprises the stored elastic energy, the regularized fracture energy,
and the work of the external loads. Here $G_{\mathrm{f}}$ is the fracture
energy, $\ell_o$ the regularization length that controls the width of the
crack band, $\alpha(d)$ the geometric crack function, and
$c_{\alpha}=4\int_{0}^{1}\sqrt{\alpha(\beta)}\,\mathrm{d}\beta$ a
normalization constant ensuring that $G_{\mathrm{f}}$ is recovered in the
sharp-crack limit $\ell_o\to0$. The strain energy density couples the two
fields through a spectral tension--compression split
\cite{miehe2010thermodynamically}, so that only tensile deformation drives
damage,
\begin{equation}\label{eq:strain-energy-split}
\psi_{s}(\boldsymbol{\varepsilon},d)
=\omega(d)\,\psi_{0}^{+}(\boldsymbol{\varepsilon})+\psi_{0}^{-}(\boldsymbol{\varepsilon}),
\qquad
\psi_{0}(\boldsymbol{\varepsilon})=\tfrac12\lambda\,\mathrm{tr}^{2}(\boldsymbol{\varepsilon})+\mu\,\mathrm{tr}\!\left(\boldsymbol{\varepsilon}^{2}\right),
\end{equation}
where $\omega(d)$ is a monotonically decreasing degradation function with
$\omega(0)=1$ and $\omega(1)=0$,
$\psi_{0}^{\pm}$ are the tensile and compressive parts of the undamaged strain
energy density $\psi_0$ obtained from the spectral decomposition of
$\boldsymbol{\varepsilon}$ \cite{miehe2010thermodynamically}, and $\lambda$,
$\mu$ are the Lam\'e constants of the undamaged material. To keep the
displacement subproblem well posed as $d\to1$, the implementation replaces
$\omega(d)$ by $\omega(d)+k_{\mathrm{res}}$ with a residual stiffness
$0<k_{\mathrm{res}}\ll1$, so that the stiffness of fully broken material does
not vanish; since $k_{\mathrm{res}}$ is constant, $\omega'(d)$ and hence the
phase-field equation are unaffected. We use $k_{\mathrm{res}}=10^{-8}$
throughout, except in the hydraulic fracture example, which follows
\citet{zhou2023hydraulic} with $(1-k_{\mathrm{res}})\,\omega(d)+k_{\mathrm{res}}$
and $k_{\mathrm{res}}=10^{-9}$. The three-dimensional examples
(\cref{ex:sent3d,ex:molar}) use the hybrid formulation of
\citet{ambati2015review}, in which the stress is degraded isotropically,
$\boldsymbol{\sigma}=\big(\omega(d)+k_{\mathrm{res}}\big)\,\partial\psi_{0}/\partial\boldsymbol{\varepsilon}$,
while the crack-driving history field is still built from the tensile part
$\psi_{0}^{+}$. This renders the displacement subproblem linear at fixed $d$,
at the price of variational consistency.
Stationarity of \cref{eq:total-energy} with respect to $\boldsymbol{u}$ and
$d$ yields the coupled governing equations
\begin{equation}\label{eq:strong-form}
\nabla\cdot\boldsymbol{\sigma}+\boldsymbol{b}^{*}=\boldsymbol{0},
\qquad
\omega'(d)\,\mathcal{H}+\frac{G_{\mathrm{f}}}{c_{\alpha}}\left[\frac{\alpha'(d)}{\ell_o}-2\ell_o\,\nabla^{2}d\right]=0
\quad\text{in }\varOmega,
\end{equation}
with the Cauchy stress
$\boldsymbol{\sigma}=\partial\psi_{s}/\partial\boldsymbol{\varepsilon}$, the
natural conditions $\boldsymbol{\sigma}\cdot\boldsymbol{n}=\boldsymbol{t}^{*}$
and $\nabla d\cdot\boldsymbol{n}=0$ on $\partial\varOmega_{N}$, and
$\boldsymbol{u}=\boldsymbol{u}^{*}$ on $\partial\varOmega_{D}$. Damage
irreversibility ($\dot{d}\ge0$, $0\le d\le1$) is enforced by driving the phase
field with a history field that records the maximum tensile energy attained
up to the current time $t$ \cite{miehe2010phase},
$\mathcal{H}(\boldsymbol{x},t)=\max_{\tau\le t}\psi_{0}^{+}\big(\boldsymbol{\varepsilon}(\boldsymbol{x},\tau)\big)$.

The pair $\{\alpha(d),\omega(d)\}$ distinguishes the two models considered in
this work. The \texttt{AT2} model \cite{Bourdin2000,miehe2010thermodynamically}
uses $\alpha(d)=d^2$ and $\omega(d)=(1-d)^2$; it is restricted to brittle
fracture and damages under any load. The phase-field cohesive zone model
\texttt{PF-CZM} \cite{wuUnifiedPhasefieldTheory2017,wuLengthScaleInsensitive2018a}
adopts $\alpha(d)=2d-d^2$ with the rational degradation function
\begin{equation}\label{eq:pfczm-deg}
\omega(d)=\frac{(1-d)^2}{(1-d)^2+a_1\,d\,P(d)},
\qquad
a_1=\frac{4}{\pi}\cdot\frac{l_{\mathrm{ch}}}{\ell_o},
\quad
P(d)=1+a_2d+a_2a_3d^2,
\end{equation}
where $l_{\mathrm{ch}}=E_0G_{\mathrm{f}}/f_{\mathrm{t}}^2$ is Irwin's
characteristic length, $E_0$ the Young's modulus, and $f_{\mathrm{t}}$ the
tensile strength. The coefficients $\{a_2,a_3\}$ calibrate a target
traction--separation law, which renders \texttt{PF-CZM} insensitive to the
length scale $\ell_o$ and applicable to both brittle and cohesive fracture.
The linear term of $\alpha(d)$ also gives \texttt{PF-CZM} a finite elastic
threshold before damage initiates. Both models are summarized in
\cref{tab:pfm}.

\begin{table}[!htb]
\small
\centering
\renewcommand\arraystretch{1.3}
\caption{Phase-field models considered in this study, with
$l_{\mathrm{ch}}=E_0G_{\mathrm{f}}/f_{\mathrm{t}}^2$ Irwin's characteristic
length. For \texttt{PF-CZM}, linear softening uses
$a_2=-\tfrac{1}{2}$, $a_3=0$, and concrete softening uses $a_2=1.3868$,
$a_3=0.9106$.}
\label{tab:pfm}
\begin{tabularx}{0.7\textwidth}{lXX}
\toprule
 & \texttt{AT2} & \texttt{PF-CZM} \\
\midrule
$\alpha(d)$ & $d^2$ & $2d-d^2$ \\[4pt]
$\omega(d)$ & $(1-d)^{2}$ & $\dfrac{(1-d)^2}{(1-d)^2+a_1d\,P(d)}$ \\[10pt]
$c_\alpha$ & $2$ & $\pi$ \\[4pt]
Length scale $\ell_o$ & $\dfrac{27}{256}\,l_{\mathrm{ch}}$ & $0<\ell_o\ll l_{\mathrm{ch}}$ \\[10pt]
Fracture type & brittle & brittle \& cohesive \\
Elastic domain & no & yes \\
\bottomrule
\end{tabularx}
\end{table}

\subsection{Realization in \texttt{Axion}}\label{sec:pff-in-axion}

In the notation of \cref{sec:framework}, the problem consists of two
subproblems, $\Phi=\{u,d\}$, with $\mathbf{z}^{u}=\mathbf{u}$ and
$\mathbf{z}^{d}=\mathbf{d}$ the nodal displacement and damage vectors. The
kernel $\mathcal{F}^{u}_{e}$ evaluates the weak form of the first equation of
\cref{eq:strong-form} with $d$ held fixed, and $\mathcal{F}^{d}_{e}$ that of
the second with $\boldsymbol{u}$ and $\mathcal{H}$ held fixed; their tangents
follow from \cref{eq:vectorized_tangent_construction}. The two models differ
only in the functions $\alpha$ and $\omega$ inside these kernels. The history
field $\mathcal{H}$ is the internal variable of the problem: it is stored at
the quadrature points and updated after each displacement solve, as in the
staggered scheme of \cite{miehe2010phase,Bourdin2000}, which
\cref{alg:staggered_phase_field} realizes for $\Phi=\{u,d\}$.

The damage subproblem is subject to the bounds $0\le d\le1$. For \texttt{AT2},
the history-driven subproblem is linear and symmetric positive definite, and
because $\mathcal{H}\ge0$ its solution lies in $[0,1]$ up to discretization
error; it is solved by a single Newton step with the CG solver of
\cref{sec:matfree}, and the projection onto $[0,1]$ only removes small
overshoots. For \texttt{PF-CZM}, neither property holds. The lower bound is
active from the first load increment, since the linear term of
$\alpha(d)=2d-d^{2}$ creates a finite elastic threshold
$\mathcal{H}_{\mathrm{th}}=f_{\mathrm{t}}^{2}/(2E_0)$ below which the
unconstrained problem would return $d<0$. Moreover, $\alpha''(d)=-2$
contributes a negative term to the damage tangent: at $d=0$ the local
curvature $\omega''(0)\,\mathcal{H}-2G_{\mathrm{f}}/(\pi\ell_o)$ is negative
for $\mathcal{H}<\mathcal{H}_{\mathrm{th}}/[2(a_1-a_2-2)]$, and together with
the rational degradation function of \cref{eq:pfczm-deg} this makes the
subproblem non-convex and its tangent indefinite. CG does not apply to such a
system, and in our tests Newton-CG iterations for the \texttt{PF-CZM} damage
field produced non-physical damage. It is therefore solved by the projected
Jacobi iteration \cref{eq:projected_jacobi}, which needs only residual
evaluations.

The octree refinement of \cref{subsec:octree_adaptivity} is well suited to
this problem, which concentrates its nonlinearity in a band of width
$\mathcal{O}(\ell_o)$ around the crack that must be resolved by elements of
size $h\lesssim\ell_o/2$. In the adaptive computations, the refinement
indicator $\eta_e$ is the maximum nodal damage of the cell, so that refinement
follows the crack band. Because damage is irreversible, refinement without
coarsening suffices.

\subsection{Extension to hydraulic fracture in a poroelastic medium}
\label{sec:hf-formulation}

Hydraulic fracture couples fluid pressure to deformation and damage, so the
discrete system carries three subproblems rather than two, and the fluid
properties themselves depend on the phase field as the material passes from
intact to fully broken. The formulation below follows
\citet{zhou2023hydraulic} and extends the two-field system of
\cref{sec:formulation} by a pore pressure $p(\boldsymbol{x},t)$, so that
$\Phi=\{u,d,p\}$. It is set out here because it is the case in which
hand-derived tangent operators become laborious: each new coupling term
contributes to several blocks of the linearized system, and none of those
blocks has to be written down in \texttt{Axion}. The energy functional retains the elastic and regularized fracture terms of
\cref{eq:total-energy} and adds the work of the in-situ stress field
$\boldsymbol{\sigma}_0$ and of the pore pressure,
\begin{equation}\label{eq:hf-energy}
\mathcal{L}(\boldsymbol{u},p,d)
=\int_{\varOmega}\left[\psi_{s}(\boldsymbol{\varepsilon},d)+\omega(d)\,\boldsymbol{\sigma}_0:\boldsymbol{\varepsilon}-\alpha_{\mathrm{B}}\,p\,(\nabla\cdot\boldsymbol{u})+\frac{G_{\mathrm{f}}}{c_{\alpha}}\left(\frac{\alpha(d)}{\ell_o}+\ell_o\,\nabla d\cdot\nabla d\right)\right]\mathrm{d}V-W_{\mathrm{ext}},
\end{equation}
where $\alpha_{\mathrm{B}}$ is the Biot coefficient and $W_{\mathrm{ext}}$ the
work of the external loads. The in-situ term measures the elastic energy
accumulated between the initial and current states and carries $\omega(d)$,
so that a fully broken region transmits no in-situ stress; without it, the
loaded boundaries would deform under the applied tractions before any
injection has occurred.

Stationarity of \cref{eq:hf-energy} with respect to $\boldsymbol{u}$ gives
equilibrium in terms of a total stress that superposes the effective, in-situ,
and pore-pressure contributions, and stationarity with respect to $d$ returns
the phase-field equation of \cref{eq:strong-form} with the history field
augmented by the work of the in-situ stress,
\begin{equation}\label{eq:hf-momentum}
\boldsymbol{\sigma}=\boldsymbol{\sigma}^{e}(\boldsymbol{\varepsilon},d)+\omega(d)\,\boldsymbol{\sigma}_0-\alpha_{\mathrm{B}}\,p\,\boldsymbol{I},
\qquad
\mathcal{H}=\max_{\tau\le t}\left[\psi_0^{+}(\boldsymbol{\varepsilon})+\boldsymbol{\sigma}_0:\boldsymbol{\varepsilon}\right]_{\tau},
\end{equation}
where $\boldsymbol{\sigma}^{e}=\partial\psi_s/\partial\boldsymbol{\varepsilon}$
is the effective stress and $\boldsymbol{I}$ the second-order identity tensor.
The pore fluid is compressible and viscous, with density $\rho$, dynamic
viscosity $\mu_{\mathrm{eff}}$, permeability $K_{\mathrm{eff}}$, and Darcy
velocity $\boldsymbol{v}=-(K_{\mathrm{eff}}/\mu_{\mathrm{eff}})\nabla p$, so
that mass conservation reads
\begin{equation}\label{eq:hf-flow}
\rho S\frac{\partial p}{\partial t}-\nabla\cdot\left(\rho\,\frac{K_{\mathrm{eff}}}{\mu_{\mathrm{eff}}}\,\nabla p\right)
=q_m-\rho\,\alpha_{\mathrm{B}}\chi_{\mathrm{r}}\frac{\partial}{\partial t}(\nabla\cdot\boldsymbol{u}),
\end{equation}
where $S$ is the storage coefficient, $q_m$ the fluid source, and the last
term the poroelastic coupling through the rate of volumetric strain. Every
fluid property follows the phase field. With a reservoir region
($d\le c_1$), a fractured region ($d\ge c_2$), and indicators
$\chi_{\mathrm{f}}=(d-c_1)/(c_2-c_1)$ and $\chi_{\mathrm{r}}=1-\chi_{\mathrm{f}}$
interpolating between them, any property
$\xi\in\{\rho,c,K,\mu,\alpha_{\mathrm{B}}\}$, where $c$ denotes the fluid
compressibility and $\mu$ here the fluid viscosity, is evaluated as
$\xi=\xi_{\mathrm{r}}\chi_{\mathrm{r}}+\xi_{\mathrm{f}}\chi_{\mathrm{f}}$, so
that the crack acts as a high-conductivity channel once the damage exceeds the
threshold.

In \texttt{Axion} the pressure enters as a third subproblem, $\phi=p$. The
three subproblems are advanced by \cref{alg:staggered_phase_field} with
$\Phi=\{u,d,p\}$, and \cref{eq:hf-flow} is discretized in time by backward
differences, so that the pressure subproblem has the same algebraic structure
as the others. What does not appear in this description is the
linearization: the residuals and all coupling terms are written once as the
weak forms of \cref{eq:hf-energy,eq:hf-momentum,eq:hf-flow}, and every block
of the tangent, including the pressure--displacement coupling and the
derivatives of the interpolated properties with respect to $d$, follows from
the same forward-mode differentiation used for displacement and damage.
Extending an \texttt{Axion} model by a physical field is thus a matter of
writing the weak form, not of differentiating it.

\section{Numerical examples}\label{sec:examples}

This section assesses \texttt{Axion} on two classes of problem. The first is
the canonical mode-I fracture benchmark in its two- and three-dimensional forms,
used to verify the framework and to quantify its computational performance and
parallel scalability (\cref{sec:benchmarks}). The second is a set of
practical-scale applications that exercise the framework rather than the solver
alone: a coupled hydro-poroelastic problem, fracture of a premolar tooth at
millimeter scale, and an industrial-scale gravity dam (\cref{sec:largescale}).
\Cref{tab:examples_overview} summarizes the examples, and
\cref{tab:cost} collects the problem sizes and wall-clock times of all of them. All computations are performed in double precision on NVIDIA RTX Pro 6000
Blackwell GPUs, each with $96\,\text{GB}$ of memory and a memory bandwidth of
$1.8\,\text{TB/s}$; the double-precision throughput of this GPU is $1/64$ of
its single-precision throughput \cite{nvidia2025rtxpro}. Double precision is
used because the residual stiffness $k_{\mathrm{res}}=10^{-8}$ gives a
stiffness contrast of $10^{8}$ between intact and fully damaged material,
which exceeds the seven significant digits of single precision. \texttt{Axion}
uses \texttt{JAX}~0.10.2 with CUDA~12. Where a CPU reference
is reported, it is a \texttt{FEniCSx} implementation (\texttt{DOLFINx}
version~0.10.0) run on an Apple M4 Max CPU, with bilinear
quadrilateral or trilinear hexahedral elements, a sparse direct solver (MUMPS)
in two dimensions and GMRES with an incomplete-LU preconditioner in three for
the displacement, CG with an algebraic multigrid preconditioner for the damage,
Newton tolerances of $10^{-9}$ and $10^{-6}$ on the displacement and damage
updates, and the same staggered scheme as \texttt{Axion}. The staggered
tolerance is also the same in two dimensions, $10^{-5}$; in three dimensions
\texttt{Axion} uses $10^{-3}$ and the reference $10^{-5}$.
Reported times are wall-clock times for the complete loading history and
include compilation, except for the three-dimensional benchmark and scaling
runs, in which the kernels are compiled before the timed solve.

\begin{table}[htbp]
\centering
\small
\setlength{\tabcolsep}{5pt}
\renewcommand{\arraystretch}{1.15}
\begin{threeparttable}
\caption{Overview of the numerical examples: the number of degrees of freedom
(DoF), the total wall-clock time of \texttt{Axion} for the full loading
history on one GPU, and the capability of the framework that each example
tests. For the two benchmarks, the DoF and the time refer to the uniform mesh.
Schematics show only the domain, loading, and supports; dimensions are given
in the corresponding sections.}
\label{tab:examples_overview}
\begin{tabularx}{\textwidth}{@{}>{\raggedright\arraybackslash}m{2.4cm}
                >{\centering\arraybackslash}m{2.9cm}
                >{\centering\arraybackslash}m{1.9cm}
                >{\centering\arraybackslash}m{1.9cm}
                >{\centering\arraybackslash}m{2.9cm}
                >{\raggedright\arraybackslash}m{2.7cm}@{}}
\toprule
\textbf{Example} & \textbf{Schematic} & \textbf{DoF} & \textbf{Time}
 & \textbf{Result} & \textbf{Tests} \\
 & & [$\times10^{5}$] & (1 GPU) & & \\
\midrule
\multicolumn{6}{@{}l}{\textit{Benchmark problems}} \\[2pt]
2D mode I (\texttt{AT2})\newline (Sec.~\ref{ex:sent}) &
\includegraphics[width=\linewidth]{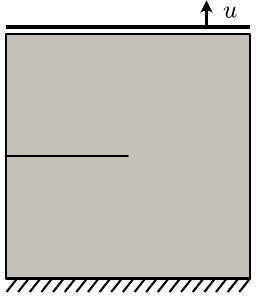} &
$0.308$ & $4.55$ min &
\includegraphics[width=\linewidth]{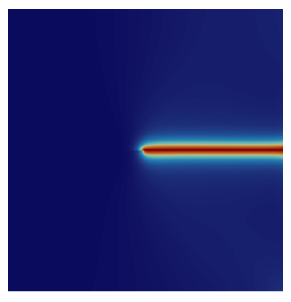} &
accuracy; uniform and adaptive meshes \\[3pt]
3D mode I (\texttt{AT2})\newline (Sec.~\ref{ex:sent3d}) &
\includegraphics[width=\linewidth]{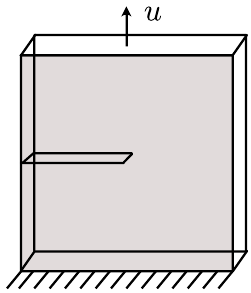} &
$14.5$ & $57.8$ min &
\includegraphics[width=\linewidth]{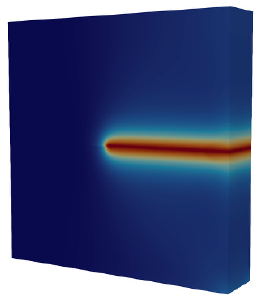} &
3D kernels; performance; parallel scaling \\
\midrule
\multicolumn{6}{@{}l}{\textit{Practical-scale problems}} \\[2pt]
Hydraulic fracture\newline (Sec.~\ref{ex:hydraulic}) &
\includegraphics[width=\linewidth]{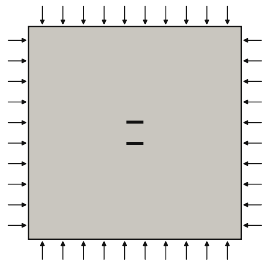} &
$3.80$ & $97$ min &
\includegraphics[width=\linewidth]{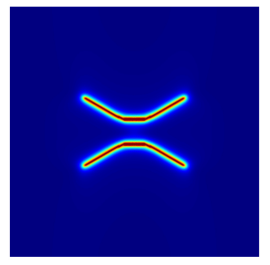} &
additional physical field (pore pressure) \\[3pt]
3D premolar (\texttt{PF-CZM})\newline (Sec.~\ref{ex:molar}) &
\includegraphics[height=3.4cm]{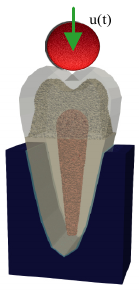} &
$6.87$ & $335$ min &
\includegraphics[height=3.4cm]{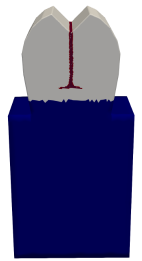} &
six materials; contact; unstructured tetrahedra \\[3pt]
Koyna dam (\texttt{AT2})\newline (Sec.~\ref{ex:koyna}) &
\includegraphics[height=3.4cm]{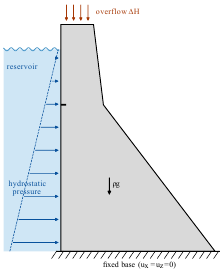} &
$960$ & $390$ min &
\includegraphics[height=3.4cm]{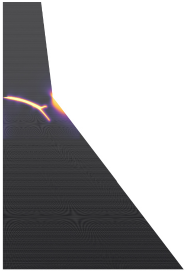} &
problem size \\
\bottomrule
\end{tabularx}
\end{threeparttable}
\end{table}

\begin{table}[htbp]
\centering
\small
\setlength{\tabcolsep}{5pt}
\renewcommand{\arraystretch}{1.15}
\begin{threeparttable}
\caption{Problem size and total wall-clock time for the full loading history
of every example. The time ratio is computed within each row as
$t_{\mathrm{CPU}}/t_{\mathrm{GPU}}$, where $t_{\mathrm{CPU}}$ is the
single-CPU \texttt{FEniCSx} time and $t_{\mathrm{GPU}}$ the single-GPU
\texttt{Axion} time, from the tabulated values and rounded to one decimal
place; the multi-GPU times do not enter it. It is a speedup at identical
discretization only for the uniform two-dimensional case, because the two
codes use different meshes in the other two rows (notes a, b, and e).}
\label{tab:cost}
\begin{tabularx}{\textwidth}{@{}>{\raggedright\arraybackslash}X l c c c c c@{}}
\toprule
\textbf{Example} & \textbf{Mesh} & \textbf{DoF} & \textbf{\texttt{FEniCSx}}
 & \multicolumn{2}{c}{\textbf{\texttt{Axion}}} & \textbf{Time ratio} \\
 & & [$\times10^{5}$] & (1 CPU) & (1 GPU) & (multi-GPU) & \\
\midrule
\multicolumn{7}{@{}l}{\textit{Benchmark problems}} \\[2pt]
2D mode I (\texttt{AT2}) & uniform  & $0.308$           & $172.23$ min         & $4.55$ min & --                  & $37.9\times$ \\
                         & adaptive & $0.113$\tnote{a}  & $7.43$ min           & $2.41$ min & --                  & $3.1\times$ \\
3D mode I (\texttt{AT2}) & uniform  & $14.5$            & $324.2$ min\tnote{b} & $57.8$ min & $15.7$ min (4 GPUs) & $5.6\times$\tnote{e} \\
\midrule
\multicolumn{7}{@{}l}{\textit{Practical-scale problems}} \\[2pt]
Hydraulic fracture            & unstructured & $3.80$ & -- & $97$ min\tnote{c}  & --                 & -- \\
                              &              &        &    & $319$ min\tnote{d} &                    &    \\
3D premolar (\texttt{PF-CZM}) & unstructured & $6.87$ & -- & $335$ min          & --                 & -- \\
Koyna dam (\texttt{AT2})      & uniform      & $960$  & -- & $390$ min          & $190$ min (2 GPUs) & -- \\
\bottomrule
\end{tabularx}
\begin{tablenotes}[flushleft]
\footnotesize
\item[a] Final count for \texttt{Axion}; the adaptive \texttt{FEniCSx} mesh ends at $0.120\times10^{5}$ DoFs.
\item[b] Adaptive \texttt{FEniCSx} computation reaching the same minimum element size, $h=\ell_o/4$, with $5.05\times10^{5}$ DoFs.
\item[c] Two parallel notches, $50$ time steps.
\item[d] Single vertical notch, $280$ time steps.
\item[e] Uniform \texttt{Axion} mesh against the adaptive \texttt{FEniCSx} mesh of note b; not a speedup at identical discretization.
\end{tablenotes}
\end{threeparttable}
\end{table}
\subsection{Benchmark: single-edge-notched tension}\label{sec:benchmarks}

\begin{figure}[!htb]
\centering
\includegraphics[width=\textwidth]{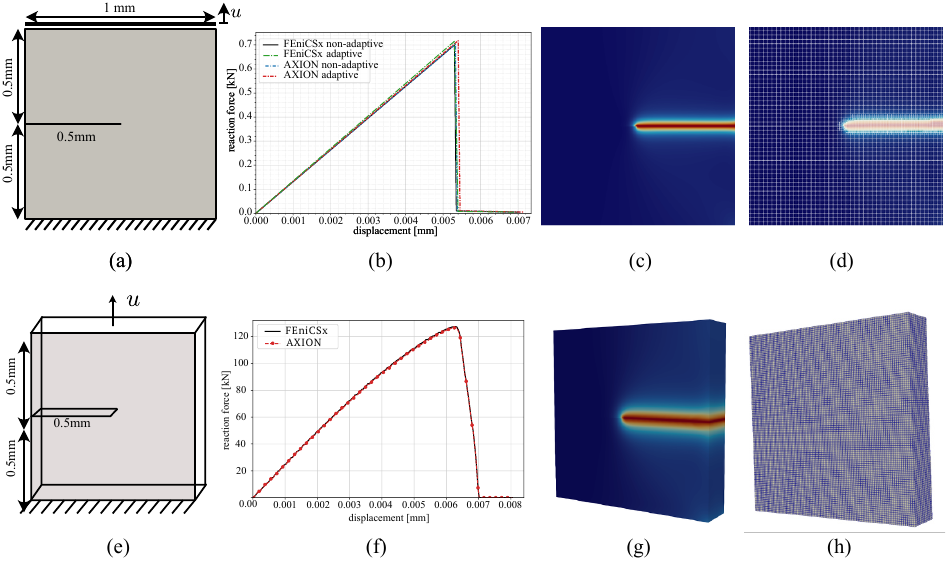}
\caption{Single-edge-notched tension, \texttt{AT2} model. Top row, two
dimensions: (a)~geometry, loading, and boundary conditions (dimensions in mm);
(b)~force--displacement response of \texttt{Axion} and \texttt{FEniCSx} in the
uniform and adaptive configurations; (c)~final damage field $d$ on the uniform
mesh at $h = \ell_o/3$; (d)~final damage field with the adaptively refined mesh
superposed. Bottom row, three dimensions: (e)~geometry, loading, and boundary
conditions; (f)~force--displacement response of \texttt{Axion} and the adaptive
\texttt{FEniCSx} reference; (g)~final damage field, showing a planar crack
whose front remains straight through the thickness; (h)~uniform hexahedral
mesh at $h = \ell_o/4$ with $1.45\times10^{6}$ DoFs.}
\label{fig:sent}
\end{figure}

The single-edge-notched tension (SENT) specimen is the canonical mode-I
benchmark for phase-field fracture and is treated here in two and three
dimensions (\cref{fig:sent}). The two-dimensional case establishes accuracy
against reference solutions and against a \texttt{FEniCSx} implementation in
both uniform and adaptive configurations. The three-dimensional case is the
principal performance study and carries the strong- and weak-scaling
investigations. Both use the \texttt{AT2} model with the material parameters of
\citet{miehe2010thermodynamically}, $E_0 = 210\,\text{kN/mm}^2$, $\nu = 0.3$,
and $G_{\mathrm{f}} = 2.7\times10^{-3}\,\text{kN/mm}$, and the regularization
length $\ell_o = 0.03\,\text{mm}$.

\subsubsection{Two-dimensional single-edge-notched tension}\label{ex:sent}

The geometry and boundary conditions are given in \cref{fig:sent}(a): a unit
square of side $L = 1\,\text{mm}$ with a horizontal pre-crack of length $L/2$
from the left edge to the center, introduced as a geometric slit by duplicating
the nodes on $\{y = L/2,\, x < L/2\}$. The bottom edge is clamped and a
vertical displacement is applied to the top edge in $700$ increments of
$\Delta\bar{u} = 1\times10^{-5}\,\text{mm}$ under plane strain. Two discretization strategies are compared, both run by \texttt{Axion} under
identical loading. The \emph{adaptive} configuration starts from a coarse $48\times48$
mesh and refines around the evolving crack with the octree scheme of
\cref{subsec:octree_adaptivity}, marking cells whose maximum nodal damage
exceeds $0.5$ and splitting them while their size exceeds $\ell_o/3$. From
the initial size of $1/48\,\text{mm}$ this takes two levels of refinement, so
that the smallest elements have $h_{\min}=1/192\,\text{mm}\approx\ell_o/5$,
about half the element size of the uniform mesh. The adaptive mesh has
$11\,334$ DoFs at the end of the simulation (\cref{fig:sent}(d)).
Earlier adaptive schemes for phase-field fracture by some of the authors are
described in \cite{krishnan2022adaptive,gupta2023amr}.

The crack propagates horizontally from the notch tip in pure mode~I, the damage
band symmetric about the crack plane and of width set by the regularization
length (\cref{fig:sent}(c)); on the adaptive mesh it is indistinguishable from
its uniform-mesh counterpart (\cref{fig:sent}(d)). The force--displacement
response, \cref{fig:sent}(b), is the one established for this benchmark
\cite{miehe2010thermodynamically,wuLengthScaleInsensitive2018a}: a linear elastic branch
terminated by an abrupt loss of load-carrying capacity, with the two
configurations coinciding throughout.

The two configurations differ in computational cost (\cref{tab:cost}). On
the uniform mesh \texttt{Axion} completes the $700$-increment history in
$4.55\,\text{min}$ against $172.23\,\text{min}$ for the single-CPU
\texttt{FEniCSx} reference, a speedup of $37.9$; in the adaptive configuration
it retains a factor of $3.1$. The smaller factor is attributed to the size of
the adaptive meshes, which have $2.7$ times fewer unknowns than the uniform
mesh, and to the cost of adaptation: every increment performs eight
refine-balance-transfer sweeps, each followed by a new staggered solve,
whether or not a cell is marked. The adaptive computation therefore takes
$0.21\,\text{s}$ per increment against $0.39\,\text{s}$ on the uniform mesh,
a reduction by a factor of $1.9$ for $2.7$ times fewer unknowns. The advantage of the GPU is
therefore larger for the larger problem. As a consequence, the uniform
\texttt{Axion} solve, at $4.55\,\text{min}$, is faster than the adaptive
\texttt{FEniCSx} solve, at $7.43\,\text{min}$, while carrying $2.6$ times as
many unknowns. Uniform refinement is therefore a practical option for
problems of this size when the solution should not depend on an error
indicator, a marking strategy, or refinement thresholds. The adaptive
\texttt{Axion} computation remains faster than the uniform one ($2.41$ against
$4.55\,\text{min}$), so that the two approaches are complementary.

\subsubsection{Three-dimensional single-edge-notched tension}\label{ex:sent3d}

The three-dimensional specimen, shown in \cref{fig:sent}(e), extrudes the
two-dimensional configuration into a cuboid of side $L = 1\,\text{mm}$ and
thickness $B = 0.1\,\text{mm}$ with a through-thickness pre-crack of length
$L/2$ at mid-height. The bottom face is clamped, the top face is displaced
vertically in $160$ increments of $\Delta\bar{u} = 5\times10^{-5}\,\text{mm}$, and
the domain is discretized with a uniform grid of $150\times150\times15$
trilinear hexahedra (\cref{fig:sent}(h)), with an element size of
$h = 1/150\,\text{mm}\approx\ell_o/4.5$ in the plane of the specimen and $2h$
through the thickness. With the nodes duplicated along the pre-crack, the mesh
has $366\,016$ nodes and $1.46\times10^{6}$ DoFs. A uniform
discretization of this size was not run on a single CPU; the reference is
instead the adaptive \texttt{FEniCSx} computation, which reaches the same
minimum element size along the crack path with $5.05\times10^{5}$ DoFs.

The \texttt{AT2} model produces a planar mode-I crack whose front stays
straight through the thickness (\cref{fig:sent}(g)). The response, \cref{fig:sent}(f), differs from its two-dimensional
counterpart: the pre-peak branch curves over as damage accumulates ahead of
the notch, and the load then falls steeply but over a finite displacement
range rather than discontinuously. The two computations differ in the energy
split (spectral in two dimensions, hybrid in three), in the load increment
($10^{-5}$ and $10^{-4}\,\text{mm}$), and in the element size ($\ell_o/3$ and
$\ell_o/4$), and these differences, rather than the dimension, are the likely
causes of the gradual loss of stiffness before the peak. \texttt{Axion} predicts a peak load of
$126.6\,\text{N}$ against $127.4\,\text{N}$ for the reference, a difference of
$0.63$\%, with the curves indistinguishable throughout. Because this agreement holds between two solvers that use the same trilinear
hexahedral element on different meshes, it verifies the three-dimensional
element residuals. The tangents affect the convergence of
Newton's method but not the converged solution; they are verified separately
in \cref{sec:adverify}. \texttt{Axion} completes the uniform solve in $57.8\,\text{min}$,
against $324.2\,\text{min}$ for the adaptive reference (\cref{tab:cost}), on a
problem $2.87$ times larger.

\subsubsection{Parallel scaling}\label{sec:scaling}

\begin{figure}[!htb]
\centering
\includegraphics[width=\textwidth]{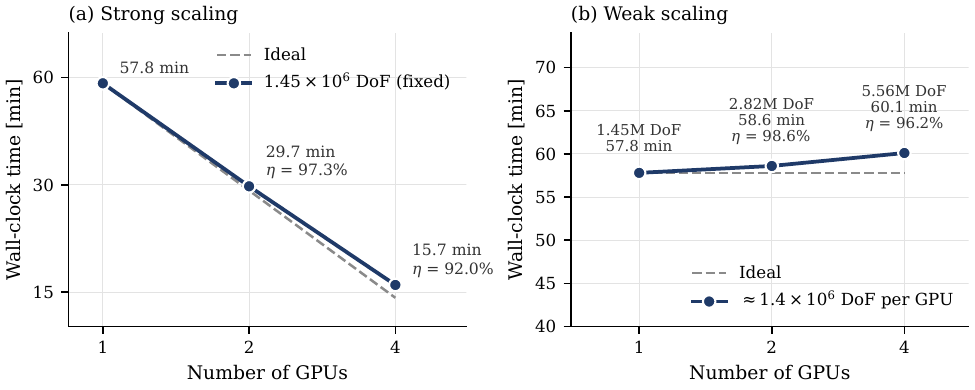}
\caption{Parallel scaling of \texttt{Axion} on the three-dimensional
single-edge-notched tension problem (\texttt{AT2}). (a)~Strong scaling at a
fixed size of $1.45\times10^{6}$ DoFs; (b)~weak scaling at approximately
$1.4\times10^{6}$ DoFs per GPU. The dashed lines indicate ideal scaling, and
$\eta$ denotes the parallel efficiency relative to the single-GPU run.}
\label{fig:scaling}
\end{figure}

Strong scaling is assessed on the uniform mesh of \cref{ex:sent3d}, partitioned by METIS into one subdomain per GPU, at a fixed size of $1.45\times10^{6}$ DoFs, and weak scaling by holding the number of DoFs
per device near $1.4\times10^{6}$ while the total problem grows to
$2.82\times10^{6}$ on two GPUs and $5.56\times10^{6}$ on four (\cref{fig:scaling}).
The additional resolution in the weak-scaling runs is placed in the
through-thickness and far-field directions, so the crack band is discretized
identically in all cases, and the predicted peak load is $126.6\,\text{N}$ in
every run. The distributed solver therefore reproduces the single-device result. 
Under strong scaling the wall-clock time falls from $57.8$ to $29.7$ and
$15.7\,\text{min}$, a four-device speedup of $3.68$ at an efficiency of
$92.0$\%. The loss of efficiency is consistent with the runtime model of
\cref{eq:distributed_time_model}: local element work scales as $1/D$, whereas
the interface exchange and the global reductions of the Krylov solver do not.
Under weak scaling the time rises only from $57.8$ to $60.1\,\text{min}$ while
the problem grows by a factor of $3.8$, an efficiency of $96.2$\%; since the
load per device falls by about $4$\% over this range, the efficiency is
optimistic by a comparable margin. Peak memory per device falls from $7.14$ to
$3.15\,\text{GB}$ under strong scaling and stays between $7.02$ and
$7.14\,\text{GB}$ under weak scaling, as \cref{eq:distributed_memory_model}
predicts, so that, within the range tested, the accessible problem size grows in proportion to the number of devices.

\subsubsection{Verification of the automatically differentiated derivatives}
\label{sec:adverify}

Two derivatives are verified: the element tangents, which are obtained by
forward-mode differentiation of the element residuals, and the derivative of
a computed output with respect to a model parameter. Both tests use the
single-edge-notched specimen of \cref{ex:sent} with the \texttt{AT2} model,
$150\times150$ elements and $\ell_o=0.02\,\text{mm}$, loaded to
$\bar{u}=5\times10^{-3}\,\text{mm}$ in $50$ increments of
$10^{-4}\,\text{mm}$, at which the largest nodal damage is $0.27$. The
staggered tolerance is $10^{-10}$, the Newton tolerance on the residual norm
is $10^{-8}$, and the relative CG tolerance is $10^{-10}$ unless stated
otherwise.

The tangents are verified by the Taylor remainder test
\cite{farrell2013automated}. For the converged state $\mathbf{z}$ at
$\bar{u}$ and a random direction $\delta\mathbf{z}$ that vanishes on the
Dirichlet unknowns and is scaled so that
$\|\delta\mathbf{z}\|_\infty=\|\mathbf{z}\|_\infty$, the remainders
\begin{equation}
r_0(h)=\big\|\mathbf{R}(\mathbf{z}+h\,\delta\mathbf{z})-\mathbf{R}(\mathbf{z})\big\|_2,
\qquad
r_1(h)=\big\|\mathbf{R}(\mathbf{z}+h\,\delta\mathbf{z})-\mathbf{R}(\mathbf{z})-h\,\mathbf{K}(\mathbf{z})\,\delta\mathbf{z}\big\|_2
\label{eq:taylor_test}
\end{equation}
are evaluated on the free unknowns for $h=10^{-1},\ldots,10^{-8}$. Where the
residual is smooth, a consistent tangent gives a first-order decrease of
$r_0$ and a second-order decrease of $r_1$ until round-off dominates. The
displacement residual is only piecewise smooth, because the spectral split
changes branch where a principal strain or the trace of the strain changes
sign, and the perturbation moves some quadrature points across a branch. The
remainder $r_1$ is therefore also evaluated on the unknowns whose elements
all remain on the same branch between $\mathbf{z}$ and
$\mathbf{z}+h\,\delta\mathbf{z}$. This restricted set contains $93\%$ of the
free unknowns at $h=10^{-4}$ and $99.9\%$ at $h=10^{-7}$, but less than
$4\%$ for $h\ge10^{-2}$, where the restricted remainder is not meaningful.

The results are listed in \cref{tab:adverify_taylor}. The order of $r_0$ is
$1.00$ at every $h$. On the restricted set, the order of $r_1$ is $1.98$,
$1.99$, $2.00$ and $2.00$ at $h=10^{-4},\ldots,10^{-7}$, and round-off
dominates below $h\approx10^{-7}$. Over all free unknowns the order of $r_1$
varies between $1.2$ and $3.6$, because the set of elements that change
branch depends on $h$. The \texttt{AT2} damage residual is linear in the
damage, so its remainder $r_1$ is at round-off level for every $h$ and the
test is uninformative. It is applied instead to the \texttt{PF-CZM} damage
residual, at the state reached by that model under the same loading and with
$\|\delta\mathbf{z}\|_\infty=1$; the order of $r_1$ is $2.03$ at
$h=10^{-2}$ and $2.00$ at $h=10^{-3},\ldots,10^{-7}$.

\begin{table}[htbp]
\centering
\small
\caption{Taylor remainder test at $\bar{u}=5\times10^{-3}\,\text{mm}$. The
order at step $h$ is $\log_{10}[r(10h)/r(h)]$. The restricted $r_1$ includes
only the unknowns whose elements remain on the same tension/compression
branch of the spectral split between $\mathbf{z}$ and
$\mathbf{z}+h\,\delta\mathbf{z}$.}
\label{tab:adverify_taylor}
\begin{tabular}{@{}l cc cc cc cc@{}}
\toprule
 & \multicolumn{6}{c}{\textbf{Displacement residual (\texttt{AT2})}}
 & \multicolumn{2}{c}{\textbf{Damage residual (\texttt{PF-CZM})}} \\
\cmidrule(lr){2-7}\cmidrule(l){8-9}
$h$ & $r_0$ & order & $r_1$ & order & $r_1$, same branch & order & $r_1$ & order \\
\midrule
$10^{-1}$ & $1.50\times10^{4}$  & --     & $3.01\times10^{2}$   & --     & $1.07\times10^{0}$   & --     & $2.02\times10^{-3}$  & --     \\
$10^{-2}$ & $1.49\times10^{3}$  & $1.00$ & $1.06\times10^{1}$   & $1.45$ & $8.44\times10^{-1}$  & $0.10$ & $1.90\times10^{-5}$  & $2.03$ \\
$10^{-3}$ & $1.49\times10^{2}$  & $1.00$ & $1.36\times10^{-1}$  & $1.89$ & $4.78\times10^{-2}$  & $1.25$ & $1.90\times10^{-7}$  & $2.00$ \\
$10^{-4}$ & $1.49\times10^{1}$  & $1.00$ & $8.03\times10^{-3}$  & $1.23$ & $4.96\times10^{-4}$  & $1.98$ & $1.90\times10^{-9}$  & $2.00$ \\
$10^{-5}$ & $1.49\times10^{0}$  & $1.00$ & $3.50\times10^{-4}$  & $1.36$ & $5.04\times10^{-6}$  & $1.99$ & $1.90\times10^{-11}$ & $2.00$ \\
$10^{-6}$ & $1.49\times10^{-1}$ & $1.00$ & $9.15\times10^{-8}$  & $3.58$ & $5.04\times10^{-8}$  & $2.00$ & $1.90\times10^{-13}$ & $2.00$ \\
$10^{-7}$ & $1.49\times10^{-2}$ & $1.00$ & $3.66\times10^{-9}$  & $1.40$ & $5.04\times10^{-10}$ & $2.00$ & $1.90\times10^{-15}$ & $2.00$ \\
$10^{-8}$ & $1.49\times10^{-3}$ & $1.00$ & $1.86\times10^{-11}$ & $2.29$ & $1.86\times10^{-11}$ & $1.43$ & $5.95\times10^{-17}$ & $1.50$ \\
\bottomrule
\end{tabular}
\end{table}

\begin{figure}[htbp]
    \centering
    \includegraphics[width=0.85\textwidth]{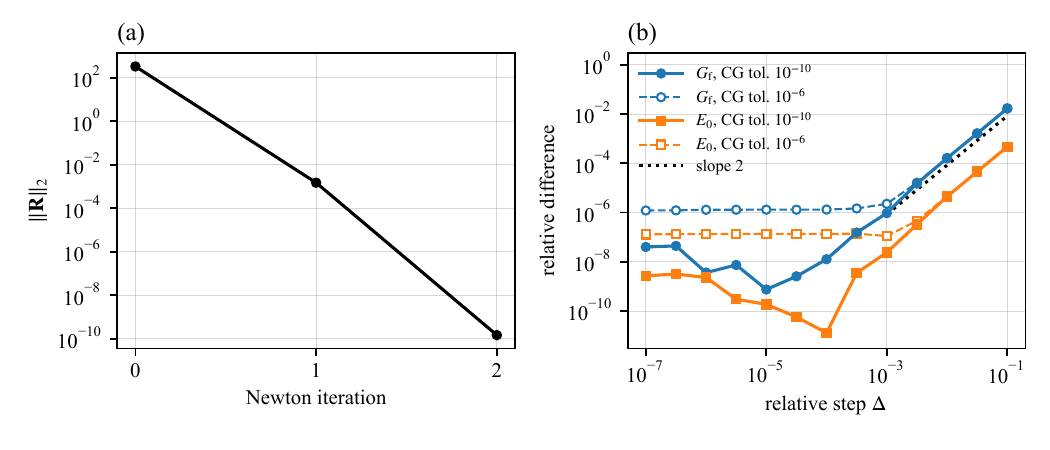}
    \caption{Verification of the derivatives for the two-dimensional
    benchmark (\texttt{AT2}). (a) Residual norm of Newton's method on the
    displacement subproblem of increment $50$. (b) Relative difference
    between the forward-mode derivative of the reaction force and the central
    difference, as a function of the relative step $\Delta$, for CG
    tolerances of $10^{-10}$ (filled symbols) and $10^{-6}$ (open symbols).}
    \label{fig:adverify}
\end{figure}

Newton's method on the displacement subproblem of increment $50$, with the
linear systems solved to a relative tolerance of $10^{-13}$, reduces the
residual norm from $3.3\times10^{2}$ to $1.5\times10^{-3}$ and then to
$1.5\times10^{-10}$ (\cref{fig:adverify}(a)). The reduction per iteration
improves from $4.5\times10^{-6}$ to $9.8\times10^{-8}$, which indicates
superlinear convergence; two iterations are too few to determine its order.

The sensitivity of the reaction force $F$ at $\bar{u}$ with respect to the
fracture energy $G_{\mathrm{f}}$ and Young's modulus $E_0$ is computed by
forward-mode differentiation of the loading history. The tangents are
propagated through the load increments and the staggered iterations, and each
nonlinear solve is differentiated implicitly at its converged state, by one
additional linear solve with the converged tangent matrix. The result is
compared with the central difference
$[F(p(1+\Delta))-F(p(1-\Delta))]/(2p\Delta)$, where $p$ is the parameter and
$\Delta$ the relative step; the two perturbed histories are computed by the
same algorithm. In every increment the damage remains below the upper bound
at all nodes, and at $\bar{u}$ no node is held at the lower bound. The
history field, which is updated after every displacement solve, exceeds the
tensile energy of the converged state by at most $3.1\times10^{-4}$ of the
maximum tensile energy (\cref{sec:diffmap}).

The normalized sensitivities $(p/F)\,\partial F/\partial p$ are $0.906$ for
$E_0$ and $0.094$ for $G_{\mathrm{f}}$; they sum to one because scaling $E_0$
and $G_{\mathrm{f}}$ by a common factor scales $F$ by that factor in the
\texttt{AT2} model. For both parameters, the relative difference between the
forward-mode derivative and the central difference decreases at second order
in $\Delta$ for $\Delta\ge3\times10^{-3}$ and is below $10^{-9}$ at
$\Delta=10^{-5}$, where it is $2\times10^{-10}$ for $E_0$ and
$8\times10^{-10}$ for $G_{\mathrm{f}}$ (\cref{fig:adverify}(b)). For
smaller steps the central difference is limited by round-off and the solver
tolerances, and the difference rises again, to $3\times10^{-9}$ for $E_0$
and $4\times10^{-8}$ for $G_{\mathrm{f}}$.

With a CG tolerance of $10^{-6}$, the difference follows the same
second-order decrease for $\Delta\ge3\times10^{-3}$ but then levels off, at
$1.4\times10^{-7}$ for $E_0$ and $1.3\times10^{-6}$ for $G_{\mathrm{f}}$ for
$\Delta\le3\times10^{-4}$ (open symbols in \cref{fig:adverify}(b)). This
level is the error of the forward-mode derivative and not of the central
difference. The central differences computed with the two CG tolerances
agree to within $3\times10^{-10}$ at $\Delta=10^{-5}$, whereas the
forward-mode derivatives computed with the two tolerances differ by
$1.4\times10^{-7}$ for $E_0$ and $1.3\times10^{-6}$ for $G_{\mathrm{f}}$,
which are the two levels. The central difference is insensitive to the CG
tolerance because the Newton iteration corrects the error of an inexact
linear solve, whereas each tangent is obtained from a single linear solve
and inherits its error.

For the \texttt{PF-CZM} model at $\bar{u}=10^{-4}\,\text{mm}$, below the
damage threshold, the damage field is zero and the reaction force is
proportional to $E_0$, so that $\partial F/\partial E_0=F/E_0$ exactly; the
forward-mode derivative reproduces this value to a relative error below
$10^{-12}$.

\subsection{Practical-scale problems}\label{sec:largescale}

The benchmarks of \cref{sec:benchmarks} establish accuracy and quantify
performance on a single canonical problem. This subsection turns to
applications that exercise the framework rather than the solver alone. Each
stresses a different aspect of \texttt{Axion}: the hydraulic fracture problem
tests whether the differentiable element formulation extends to a coupled
system with an additional physical field, the premolar tests geometric
complexity and heterogeneous material data, and the dam tests the largest problem size. None of the three is offered as a quantitative validation study; the
governing formulations are taken from the literature in each case, and the
purpose is to show that problems of this kind are solved by the same framework
without modification. The material parameters of the first two examples are
listed in Appendix~\ref{app:params}.

\subsubsection{Hydraulic fracture in a poroelastic medium}\label{ex:hydraulic}

This example applies the three-field hydro-poroelastic formulation of
\cref{sec:hf-formulation} to two configurations from
\citet{zhou2023hydraulic}, in each of which the fracture path is governed by
the in-situ stress field rather than by the notch geometry. Both share the
square domain of side $10\,\text{m}$ of that work. Notches of length
$0.8\,\text{m}$ are prescribed through an initial history field, fluid is
injected into them at $10\,\text{kg}/(\text{m}^3\cdot\text{s})$, the pressure
is held at $p = 0$ on all outer boundaries, and the time increment is
$\Delta t = 0.05\,\text{s}$. The material and fluid parameters are those of
\cref{tab:hf_params}, in which the ratio of the permeabilities of the fractured and reservoir regions is $8.3\times10^{11}$. \texttt{Axion}
discretizes the domain with an unstructured Gmsh triangulation generated on one
quadrant and reflected about both axes, which makes the mesh symmetric about
both axes without the directional bias of a Cartesian grid. The mesh has
$188\,968$ linear triangles and $95\,057$ nodes, and its longest edge of
$0.0477\,\text{m}$ gives $\ell_o/h_{\max} = 2.1$, as in the reference. With two
displacement, one pressure, and one damage unknown per node, the three fields
total $380\,228$ degrees of freedom. On a single GPU the two-notch simulation
($50$ time steps) runs in $1.62$ h and the varying-stress simulation ($280$
time steps) in $5.32$ h.

\begin{figure}[!htb]
    \centering
    \includegraphics[width=0.99\textwidth]{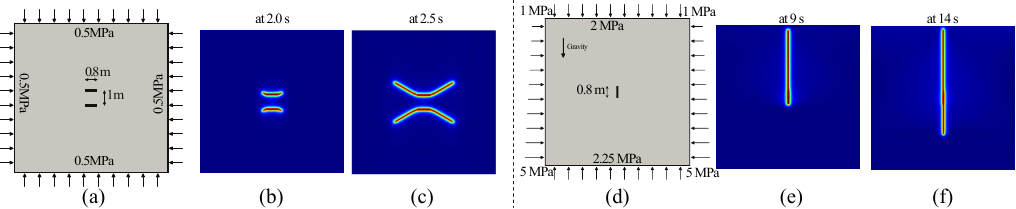}
    \caption{Hydraulic fracture in a poroelastic medium. (a)~Two parallel
    horizontal notches of length $0.8\,\text{m}$, separated vertically by
    $1\,\text{m}$, under a uniform remote stress of $0.5\,\text{MPa}$;
    (b and c)~phase field at $t = 2.0$ and $2.5\,\text{s}$, showing the
    outward deflection of the two fractures caused by stress shadowing;
    (d)~single vertical notch under gravity, with the horizontal confinement
    increasing from $1\,\text{MPa}$ at the top to $5\,\text{MPa}$ at the base;
    (e and f)~phase field at $t = 9$ and $14\,\text{s}$, showing growth from
    the upper tip alone followed by the later advance of the lower tip.}
    \label{fig:hydraulic}
\end{figure}

\paragraph{Two parallel notches under uniform in-situ stress:} Two horizontal
notches, centered in the domain and separated vertically by $1\,\text{m}$, are
pressurized simultaneously under equal remote stresses of $0.5\,\text{MPa}$
(\cref{fig:hydraulic}(a)); the injection is distributed over a band of
half-width $0.075\,\text{m}$ about each notch, a width the reference does not
specify. Under isotropic confinement a single notch would extend along its own
axis, so any deflection is caused by the neighboring fracture, whose pressure
raises the stress against which the other must open. \texttt{Axion} captures
this stress shadowing: at $t = 2.0\,\text{s}$ the four tips are already turning
away from the mid-plane (\cref{fig:hydraulic}(b)), and by $t = 2.5\,\text{s}$
they have splayed into a pattern symmetric about both axes
(\cref{fig:hydraulic}(c)) of half-length $2.14\,\text{m}$, against the
$2.5\,\text{m}$ measured from Fig.~18(b) of \citet{zhou2023hydraulic} and
consistent with the stress shadowing reported for multi-stage fracturing
\cite{sobhaniaragh2019stress}.

This symmetry is not retained automatically. The fields are therefore projected
onto the reflection symmetry of the configuration at every staggered iteration,
by averaging them over the four reflections of the mesh. Because the mesh is
built by reflecting one quadrant, the symmetric field is an exact discrete
solution and the projection changes nothing while it is followed: with and
without it the half-length agrees to three decimal places up to
$t = 1.55\,\text{s}$. At the next increment the symmetric branch loses
stability, and without the projection the antisymmetric mode seeded by
round-off grows to order unity, leaving a solution in which only one diagonal
pair of tips advances (\cref{fig:hydraulic_noproj}). Under-relaxation cannot
suppress a mode whose growth factor exceeds unity, and the staggered iteration
fails at that increment for relaxation factors of $0.25$, $0.5$ and $0.75$
alike; the projection instead removes the mode from the iteration space, and
the computation converges throughout.

\paragraph{Single vertical notch under depth-varying in-situ stress:} A single vertical notch is placed at the center of the domain in a non-uniform
in-situ stress field representative of the geological setting. Gravity acts
vertically, the horizontal stress increases linearly with depth from
$1\,\text{MPa}$ at the top to $5\,\text{MPa}$ at the base, and vertical tractions of $2\,\text{MPa}$ and $2.25\,\text{MPa}$ act on the top and bottom boundaries (\cref{fig:hydraulic}(d)). The initial-stress term of
\cref{eq:hf-energy} is required here, since without it the loaded
boundaries would deform under the applied tractions before injection begins.
The fracture stays vertical, but its two tips do not advance together, because
the horizontal stress that it must open against is lower above the notch than
below it. Growth is at first entirely upward: by $t = 9\,\text{s}$ the upper
tip has advanced about $4\,\text{m}$ and is close to the top boundary, while
the lower tip has not moved (\cref{fig:hydraulic}(e)). Only then does the lower
tip begin to advance into the more strongly confined rock, and it does so more
slowly, so that at $t = 14\,\text{s}$ the lower segment has grown by about
$2\,\text{m}$ and the upper segment remains considerably the longer
(\cref{fig:hydraulic}(f)). This upward-dominated, asymmetric propagation is
consistent with the behavior reported by \citet{zhou2023hydraulic}.

The deflection of the two interacting fractures and the order in which the
tips of the single fracture advance are not prescribed; they result from the
minimization of \cref{eq:hf-energy}, subject, in the two-notch case, to the
symmetry projection described above.

\subsubsection{Three-dimensional premolar}\label{ex:molar}

This example applies \texttt{Axion} to a biomechanical problem, the fracture of
a human tooth under occlusal loading. It combines six constituents whose
stiffnesses span more than four orders of magnitude, a thin periodontal
ligament, contact loading, and a crack that must nucleate without any notch or
initial defect.

\begin{figure}[!htb]
    \centering
    \includegraphics[width=0.85\textwidth]{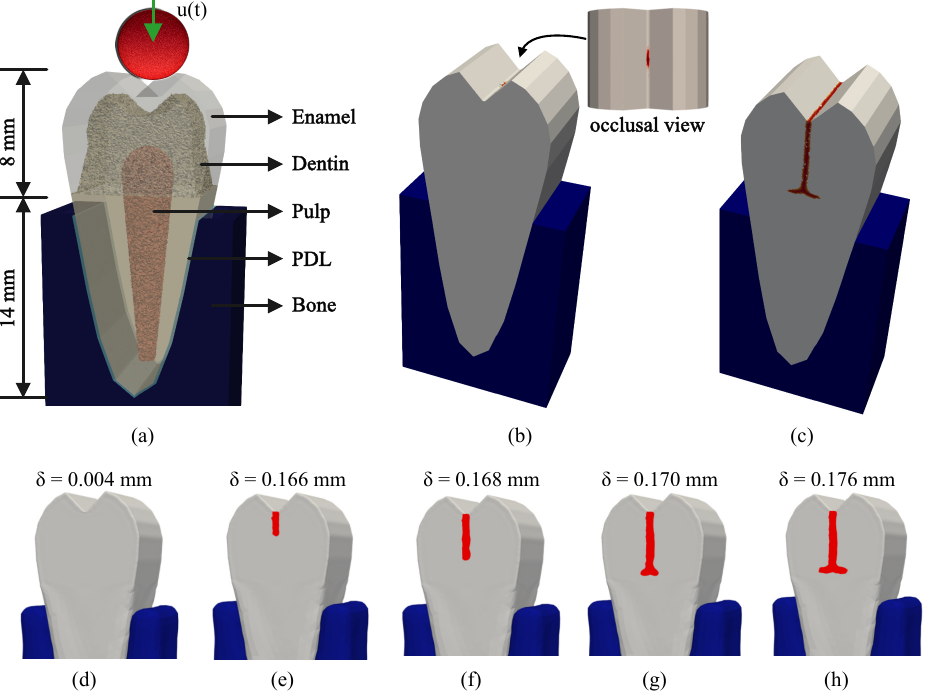}
    \caption{Three-dimensional premolar. (a)~Geometry, constituents and
    loading: a rigid sphere of $4\,\text{mm}$ diameter is driven by a
    displacement $u(t)$ into the occlusal fissure, with the base and lateral
    faces of the bone block fixed; (b)~crack initiation at the floor of the
    central fissure at an indenter displacement of
    $\delta = 0.140\,\text{mm}$, shown as the surface $d = 0.5$ (red), with the
    occlusal view in the inset; (c)~the crack at $\delta = 0.184\,\text{mm}$,
    after it has propagated through the enamel and into the dentin;
    (d)--(h)~the crack on an approximal face of the crown at
    $\delta = 0.004$, $0.166$, $0.168$, $0.170$ and $0.176\,\text{mm}$.}
    \label{fig:molar}
\end{figure}

The configuration follows the intact-tooth study of \citet{zhu2026tooth}. A
maxillary second premolar, anchored in alveolar bone through the periodontal
ligament (PDL), is loaded on its occlusal surface by a rigid sphere of
$4\,\text{mm}$ diameter driven vertically under displacement control
(\cref{fig:molar}(a)). The tooth consists of an enamel cap over a dentin core
with the pulp in its interior, and the bone block is divided into cancellous
bone and a cortical shell. An idealized geometry with the dimensions of an
average maxillary second premolar is used. The crown is $8\,\text{mm}$ high,
$9\,\text{mm}$ wide and $7\,\text{mm}$ in the mesio-distal direction, and
carries two cusps separated by a central fissure $1.7\,\text{mm}$ deep with
flanks inclined at $45^\circ$. The enamel is $1.2\,\text{mm}$ thick occlusally
and thins to $0.3\,\text{mm}$ at the cemento-enamel junction (CEJ). The root is
$14\,\text{mm}$ long, the PDL is $0.2\,\text{mm}$ thick, and the bone block
measures $12 \times 15.5 \times 7\,\text{mm}$, with its crest $1.5\,\text{mm}$
below the CEJ and a cortical shell $1\,\text{mm}$ thick. All interfaces are
perfectly bonded.

The base and the four lateral faces of the bone block are fixed. The sphere is
lowered in increments of $0.004\,\text{mm}$, refined to $0.002\,\text{mm}$ for
$\delta \in [0.164, 0.176]\,\text{mm}$ to resolve the onset of unstable crack
growth, its displacement being denoted $\delta$.
Contact is modeled
after \citet{zhu2026tooth} as a frictionless unilateral spring that acts on the
occlusal surface within $d_s = 0.3\,\text{mm}$ of the sphere, with a stiffness
of $5\times10^{3}\,\text{N/mm}^3$. Its contribution to the tangent is obtained by automatic differentiation in the same way as the bulk terms (\cref{subsec:tensorized_jit_fem}). Fracture is admitted in the
enamel and dentin of the crown, down to $1\,\text{mm}$ above the CEJ, and the
remaining constituents are elastic (\cref{tab:molar_params}). The
\texttt{PF-CZM} model \cite{wuUnifiedPhasefieldTheory2017,wuLengthScaleInsensitive2018a} is used with
$\ell_o = 0.25\,\text{mm}$, and the crack is driven by the Rankine-type measure
$\langle\bar\sigma_1\rangle^2/2E$ of the largest principal effective stress,
which vanishes under the compression beneath the indenter. The tooth, PDL and
bone are meshed as conforming volumes with linear tetrahedra, so every internal
interface coincides with element faces. The element size is
$h = 0.125\,\text{mm}$ in the crown and $0.8\,\text{mm}$ in the root and bone,
giving $687\,132$ degrees of freedom.
The full loading history completes in $5.59$ h on a single GPU.

The crack nucleates at the floor of the central fissure, midway along its
length, where the sphere wedges the two cusps apart (\cref{fig:molar}(b)). The
occlusal view in the inset shows that it starts as a short segment aligned with
the fissure, while the enamel beneath the indenter remains intact. The crack
first spreads along the fissure toward the two approximal faces and then runs
downward in the mesio-distal plane, through the enamel and into the dentin
toward the pulp. \Cref{fig:molar}(d)--(h) follow this descent on an approximal
face of the crown, which is intact at the first increment (d). At
$\delta = 0.166\,\text{mm}$ the crack has reached the face and extends a short
distance below the fissure (e), at $\delta = 0.168\,\text{mm}$ it has run most
of the way through the dentin of the crown (f), and at
$\delta = 0.170\,\text{mm}$ it arrives at the lower limit of the region in
which fracture is admitted (g). The descent through the crown therefore takes
place within about $0.004\,\text{mm}$ of indenter travel, which reflects the
brittle character of the failure. Because the crack cannot extend below that
limit, further loading spreads it sideways there, giving the foot seen in (h)
and, at $\delta = 0.184\,\text{mm}$, in \cref{fig:molar}(c).

This sequence, with damage appearing between the cusps at mid-length, growing
toward the approximal faces and through the enamel, and then propagating down
into the dentin, is the one reported for the intact premolar in the peridynamic
simulation of \citet{zhu2026tooth}. The displacement levels are not compared,
because that study uses a patient-specific geometry and a bond-stretch failure
criterion. The site and path of the crack result from the geometry, the contact
loading, and the contrast between the constituents, which enter only as
material data at the quadrature points of a statically shaped cell array.

\subsubsection{The Koyna gravity dam}\label{ex:koyna}

The Koyna dam is the standard benchmark for fracture analysis of concrete
gravity dams. It has been studied with continuum damage models
\cite{ghrib1995nonlinear}, discrete cohesive approaches
\cite{shi2013modelling}, combined XFEM and damage formulations
\cite{roth2015combined}, crack-path-field methods \cite{dias2016modeling}, and
phase-field models \cite{santillan2017dam}. The scale separation described in \cref{sec:intro} is the reason that
published analyses confine fine resolution to a region selected around the
expected crack path: the structure is $103\,\text{m}$ high, whereas the width
of the damage band, set by $\ell_o = 0.03\,\text{m}$, is more than three orders
of magnitude smaller. \texttt{Axion} instead resolves the damage band
uniformly across the entire structure, and this
example tests whether a problem of that size can be solved on a single GPU.

\begin{figure}[!htb]
    \centering
    \includegraphics[width=0.9\textwidth]{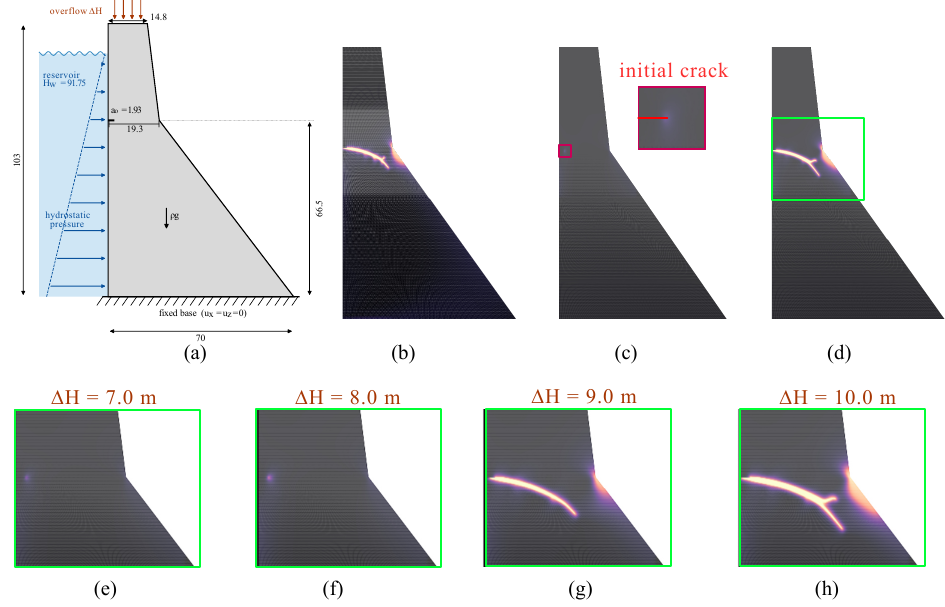}
    \caption{The Koyna gravity dam. (a)~Geometry and loading: the reservoir at
    $H_w = 91.75\,\text{m}$ with the hydrostatic pressure distribution, the
    overflow height $\Delta H$, the self-weight $\rho g$, the clamped base, and
    the initial notch of length $a_0 = 1.93\,\text{m}$ on the upstream face
    (dimensions in m). (b)~Reference computation on a locally refined mesh:
    phase field $d$ at the end of loading. (c)~Uniform mesh with
    $h = 0.015\,\text{m}$ before the onset of propagation, with an enlargement
    of the notch tip. (d)~Uniform mesh at the end of loading, showing the
    downward crack path and the secondary branch toward the change of slope on
    the downstream face; the outlined window is enlarged in (e)--(h), which
    show the phase field at overflows of $\Delta H = 7.0$, $8.0$, $9.0$ and
    $10.0\,\text{m}$.}
    \label{fig:koyna}
\end{figure}

The geometry, shown in \cref{fig:koyna}(a), follows \citet{santillan2017dam}
and the benchmark data of \citet{chopra1972koyna}. The upstream face is
vertical, the crest is $14.8\,\text{m}$ wide and the base $70\,\text{m}$ wide,
and the downstream face changes slope $36.5\,\text{m}$ below the crest, where
the local thickness is $19.3\,\text{m}$. A horizontal notch of length
$a_0 = 1.93\,\text{m}$, one tenth of the local thickness, is placed on the
upstream face at this elevation, $66.5\,\text{m}$ above the base, which earlier
studies identified as the critical location \cite{gioia1992dam}. The concrete
is homogeneous and isotropic, with $E = 25\,\text{GPa}$, $\nu = 0.2$,
$G_{\mathrm{f}} = 100\,\text{N/m}$ and density $2450\,\text{kg/m}^3$, and its
fracture is described by the \texttt{AT2} model. The base is clamped. The loads
are applied in three stages: self-weight, then hydrostatic pressure from a
reservoir at $H_w = 91.75\,\text{m}$, and finally an overflow $\Delta H$ that
is raised monotonically to $10\,\text{m}$. Plane strain is assumed, and the
water pressure inside the fracture is neglected, as in the published analyses.

The regularization length is $\ell_o = 0.03\,\text{m}$, and the mesh has a
uniform element size $h = \ell_o/2 = 0.015\,\text{m}$ over the entire cross-section, with three-node linear triangles, which gives
$9.6\times10^{7}$ degrees of freedom. The damage band is therefore resolved to the same standard everywhere, and no judgement
about where the crack will go is built into the mesh. As in the benchmarks, the kernels are compiled once (\cref{sec:compiled}).

The computed response follows the sequence established for this benchmark
\cite{santillan2017dam,gioia1992dam}, and \cref{fig:koyna}(c)--(h) show it on
the uniform mesh. As the overflow is raised the crest displacement grows
linearly and the notch remains dormant. Damage is confined to a small zone at
the notch tip (\cref{fig:koyna}(c)), which intensifies but does not extend at
$\Delta H = 7.0$ and $8.0\,\text{m}$ (\cref{fig:koyna}(e) and (f)). At an
overflow of approximately $8.3\,\text{m}$ the fracture advances rapidly under
a negligible load increment, producing a snap-through, so that at
$\Delta H = 9.0\,\text{m}$ a fully developed crack already extends across more
than half of the section (\cref{fig:koyna}(g)). Propagation then continues
along a flatter branch of the response up to the final overflow of
$10\,\text{m}$ (\cref{fig:koyna}(h)). The overflow at which the instability
occurs is not a purely geometric quantity: \citet{santillan2017dam} found it to
decrease with increasing regularization length, so $\ell_o$ acts here as an
effective material parameter rather than a numerical one.

The crack topology at the end of loading is shown in \cref{fig:koyna}(d) and
(h). The fracture leaves the notch and curves downward toward the downstream
face, driven by the compression that the hydrostatic load induces there, and at
the highest water levels a short second branch forks from it and runs toward
the change of slope. A zone of diffuse damage also forms at the change of slope
itself, where the stress concentrates; the \texttt{AT2} model has no elastic
threshold, so damage is not confined to the crack. The downward path is common
to all published analyses; the branch was identified by
\citet{jirasek1997nonlocal} and recovered by \citet{santillan2017dam}. The phase-field analysis of \citet{santillan2017dam}, carried out with their
own finite-element implementation on locally refined meshes (element size
$4$ or $8\,\text{cm}$ in a region of interest around the expected crack and
coarser three-node triangles elsewhere, between $4.0\times10^{4}$ and
$3.2\times10^{5}$ elements),
\cref{fig:koyna}(b), predicts the same downward path and the same secondary
branch, with its refined region placed where earlier studies located the crack.
The same question, restated for a
structure of this size, is whether a refined region must be placed in
advance; with a uniform mesh it need not be, so that the agreement between
\cref{fig:koyna}(b) and (d) is obtained without building any assumption about
the crack path into the discretization. Adaptive refinement also requires no
prior knowledge of the path; the octree scheme of
\cref{subsec:octree_adaptivity} is not applied here because it is restricted
to box-shaped domains.

\texttt{Axion} completes the full loading history in $6.50$ hours on a single
GPU and $3.17$ hours on two (\cref{tab:cost}), a speedup of $2.05$. The speedup slightly exceeds the ideal value, which we attribute to the single-device run operating close to the memory capacity of the GPU; the scaling study of
\cref{sec:scaling} is the more reliable guide to parallel performance. 

\section{Conclusions}\label{sec:conclusions}

This work introduced \texttt{Axion}, a differentiable, GPU-native finite
element framework written in \texttt{JAX}. The element kernels, residual
evaluation, Newton linearization, Krylov solver, bound constraints, staggered
coupling of fields, and load increments are compiled through XLA into a single
program that executes on the GPU. Consistent tangent operators are obtained by
forward-mode automatic differentiation of the element residual. Forward-mode
derivatives of the discrete solution with respect to model parameters are
obtained by propagating tangents through the load increments and staggered
iterations of the same program, with each nonlinear solve differentiated
implicitly at its converged state. The framework is extended to locally
refined meshes by an octree scheme that operates within fixed-size arrays, and
to several GPUs by a cell-based decomposition whose operator action is
identical to that of the single-device solver.

Phase-field fracture was implemented as the first physics module, with the
\texttt{AT2} and \texttt{PF-CZM} models sharing one element definition. On the
two- and three-dimensional single-edge-notched tension benchmarks,
\texttt{Axion} reproduces the reference force--displacement responses, with a
difference of $0.63\%$ in the three-dimensional peak load. The automatically
differentiated tangents pass the Taylor remainder test, with second-order
convergence of the remainder wherever the residual is smooth
(\cref{sec:adverify}). The forward-mode sensitivities of the reaction force
with respect to Young's modulus and the fracture energy agree with central
finite differences to a relative difference below $10^{-9}$ when the linear
systems are solved to a tolerance of $10^{-10}$. Their accuracy is set by
that tolerance: at a tolerance of $10^{-6}$ the difference is of order
$10^{-7}$ to $10^{-6}$. At identical discretization, the two-dimensional
benchmark completes $37.9$ times faster on one GPU than the \texttt{FEniCSx}
reference on one CPU. On one GPU, the uniformly refined computations
completed faster than the adaptive \texttt{FEniCSx} computations,
by factors of $1.6$ in two dimensions and $5.6$ in three. Uniform refinement
is therefore a practical option for problems of this size when the crack path
is not known in advance, although the adaptive \texttt{Axion} computation
remains faster than the uniform one. On four GPUs, the strong- and
weak-scaling efficiencies are $92\%$ and $96\%$. The hydraulic fracture,
premolar, and gravity dam examples were solved with the same solver and
element kernels as the benchmarks.

The present implementation has several limitations. The linear systems are
solved by Jacobi-preconditioned conjugate gradients, whose iteration count
grows with mesh refinement; the scaling results, obtained on up to four GPUs
with a fixed resolution of the damage band, should therefore not be
extrapolated to larger device counts or finer meshes without a scalable
preconditioner such as multigrid. Reverse-mode derivatives are not available
through the compiled while-loops and require implicit differentiation of each
converged increment. The parameter sensitivities were verified for the
\texttt{AT2} model on a loading history that ends before the peak load, in
which no damage unknown reaches the upper bound; where the set of active
bounds or the branch of the spectral split changes, the discrete solution is
only piecewise differentiable. The adaptive scheme is restricted to
box-shaped domains discretized by quadrilaterals or hexahedra, does not
coarsen, and has not yet been combined with the multi-GPU decomposition. The
comparisons use a single CPU.

These limitations define the next steps. Implicit differentiation of the
converged increments would provide reverse-mode derivatives for inverse
identification, optimization, and uncertainty quantification, and a scalable
preconditioner would allow larger device counts to be used. Because the solver
produces high-fidelity solutions at moderate cost, it can also be used to
generate training data for neural operators and other surrogate models, and,
because it is written in \texttt{JAX}, it can be embedded in hybrid schemes in
which the simulator is one component of a learning pipeline. The
implementation is available as described in the Code availability statement.

\newpage
\section*{Code availability}
The \texttt{Axi
on} source code and the scripts that reproduce every
example in this paper are available at
\url{https://github.com/Centrum-IntelliPhysics/AXION}.
\appendix

\section{Material parameters of the practical-scale examples}\label{app:params}

\begin{table}[!htb]
    \centering
    \caption{Hydraulic fracture: material and fluid parameters, after
    \citet{zhou2023hydraulic}. Subscripts $\mathrm{R}$ and $\mathrm{F}$ denote
    the reservoir and fractured domains.}
    \label{tab:hf_params}
    \vspace{6pt}
    \begin{tabular}{l l @{\hskip 2.5em} l l}
        \toprule
        Parameter & Value & Parameter & Value \\
        \midrule
        \addlinespace[2pt]
        Shear modulus $\mu$ [GPa]                    & $23.08$            & Lam\'e modulus $\lambda$ [GPa]                   & $34.62$ \\
        Fracture energy $G_{\mathrm{f}}$ [N/m]       & $500$              & Residual stiffness $k_{\mathrm{res}}$ [--]       & $1\times10^{-9}$ \\
        Regularization length $\ell_o$ [m]           & $0.1$              & Phase-field thresholds $c_1$, $c_2$ [--]         & $0.4$, $1.0$ \\
        Fluid density $\rho_{\mathrm{R}}, \rho_{\mathrm{F}}$ [kg/m$^3$] & $1.0\times10^{3}$ & Reservoir porosity $\varepsilon_{p\mathrm{R}}$ [--] & $0.05$ \\
        Biot coefficient $\alpha_{\mathrm{R}}$ [--]  & $0.05$             & Biot coefficient $\alpha_{\mathrm{F}}$ [--]      & $1.0$ \\
        Permeability $k_{\mathrm{R}}$ [m$^2$]        & $1\times10^{-15}$  & Permeability $k_{\mathrm{F}}$ [m$^2$]            & $8.333\times10^{-4}$ \\
        Compressibility $c_{\mathrm{R}}, c_{\mathrm{F}}$ [1/Pa] & $1\times10^{-8}$ & Viscosity $\mu_{\mathrm{R}}, \mu_{\mathrm{F}}$ [Pa$\cdot$s] & $1\times10^{-3}$ \\
        \bottomrule
    \end{tabular}
\end{table}

\begin{table}[!htb]
    \centering
    \caption{Three-dimensional premolar: material parameters. Elastic constants
    after \citet{zhu2026tooth}; fracture energies after
    \citet{yang2024multilayered}; tensile strengths after
    \citet{giannini2004uts} and \citet{sano1994dentin}.}
    \label{tab:molar_params}
    \footnotesize
    \setlength{\extrarowheight}{2pt}
    \begin{tabular}{lcccc}
        \toprule
        Constituent & $E$ [GPa] & $\nu$ [--] & $G_{\mathrm{f}}$ [N/m] & $f_t$ [MPa] \\
        \midrule
        Enamel          & $84.1$  & $0.30$ & $13.1$  & $40$ \\
        Dentin          & $18.1$  & $0.31$ & $161.8$ & $60$ \\
        Pulp            & $0.002$ & $0.45$ & --      & --   \\
        PDL             & $0.01$  & $0.45$ & --      & --   \\
        Cancellous bone & $1.37$  & $0.30$ & --      & --   \\
        Cortical bone   & $13.7$  & $0.26$ & --      & --   \\
        \midrule
        \multicolumn{5}{l}{Regularization length $\ell_o = 0.25\,\text{mm}$;
        residual stiffness $k_{\mathrm{res}} = 1\times10^{-8}$} \\
        \bottomrule
    \end{tabular}
\end{table}

\begin{figure}[H]
\centering
\includegraphics[width=1.0\textwidth]{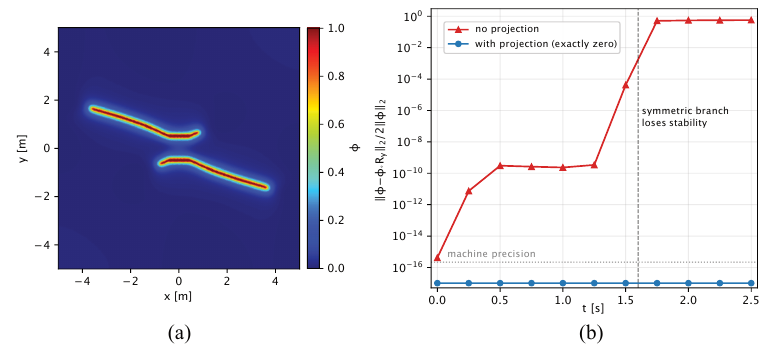}
\caption{The two-notch computation without the reflection-symmetry projection.
(a)~Phase field at $t = 2.0\,\text{s}$: only one diagonal pair of tips advances,
giving a pattern symmetric under rotation by $180^{\circ}$ but not under
reflection, to be compared with \cref{fig:hydraulic}(b). (b)~Antisymmetric
component of the phase field. The mesh is generated by reflecting one quadrant,
so the discrete problem is exactly equivariant and the projected computation
keeps this quantity at zero to machine precision.}
\label{fig:hydraulic_noproj}
\end{figure}

\bibliographystyle{unsrtnat}  
\bibliography{bibliography}

\end{document}